\documentclass[pdftex,english,12pt]{article}

\usepackage{natbib}
\usepackage{amsmath}
\usepackage{amssymb}
\usepackage{amsthm}
\usepackage{multirow}
\usepackage{graphicx}
\usepackage{colortbl}
\usepackage{authblk}
\usepackage{url}
\usepackage{subcaption, tabularx}
\usepackage[top=1in, bottom=1in, left=1in, right=1in]{geometry}

\begin{document}

\title{Activity Classification Using Smartphone Gyroscope and Accelerometer Data}

\author[]{Emily Huang\thanks{ehuang@hsph.harvard.edu} } 
\author[]{Jukka-Pekka Onnela}
\affil[]{Department of Biostatistics, Harvard University}

\maketitle

\begin{abstract}
Activities, such as walking and sitting, are commonly used in biomedical settings either as an outcome or covariate of interest. Researchers have traditionally relied on surveys to quantify activity levels of subjects in both research and clinical settings, but surveys are not objective in nature and have many known limitations, such as recall bias. Smartphones provide an opportunity for unobtrusive objective measurement of various activities in naturalistic settings, but their data tends to be noisy and needs to be analyzed with care. We explored the potential of smartphone accelerometer and gyroscope data to distinguish between five different types of activity: walking, sitting, standing, ascending stairs, and descending stairs. We conducted a study in which four participants followed a study protocol and performed a sequence of various activities with one phone in their front pocket and another phone in their back pocket. The subjects were filmed throughout, and the obtained footage was annotated to establish ground truth activity. We  applied the so-called movelet method to classify their activity. Our results demonstrate the promise of smartphones for activity detection in naturalistic settings, but they also highlight common challenges in this field of research. 

\end{abstract}

\clearpage

\section{Introduction}
\label{intro}

Many researchers have recently advocated a more substantial role for large-scale phenotyping as a route to advances in the biomedical sciences. Of the many different phenotype classes, precise capture of social, behavioral, and cognitive markers in naturalistic settings has traditionally presented special challenges to phenomics because of their temporal nature, contextual dependence, and lack of tools for measuring them objectively. The ubiquity of smartphones presents an opportunity to capture these markers in free-living settings, offering a scalable solution to the phenotyping problem. Smartphones have at least three distinct advantages compared to other approaches to social, behavioral, and cognitive phenotyping: (1) the availablility of these devices makes it possible to implement large studies without requiring additional subject instrumentation; (2) reliance on passive data makes the process unobtrusive and poses no burden on the subject, making long-term followup possible; and (3) the combination of the previous two factors make it possible, at least in principle, to obtain these markers prospectively from a cohort of interest at a low cost.

To classify different types of activity, we consider data from the tri-axial accelerometer and tri-axial gyroscope in a smartphone. These sensors output data for three orthogonal axes in the frame of reference of the phone ($x$ = from left to right, $y$ = from top to bottom, $z$ = through the phone). The gyroscope measures the angular velocity (units of radians/second) about each axis, which indicates the direction and speed that the phone is spinning about the axis. The accelerometer measures the acceleration along each axis in units of $g$ (1 $g$ = 9.81 m/s/s). % JPO: you should check this, but we should make it clear that one measures *angular velocity* and the other measures *linear acceleration*, acceleration in the direction of the axes (not around them); but please check this %Emily: I decided not to use linear acceleration since it seems that people refer to linear acceleration as removing the effect of gravity.
The sampling rates of these sensors can vary based on phone types and the mode the phone is in. In iPhones, for example, accelerometer and gyroscope data are collected at a frequency of 10 Hz, i.e., 10 samples per second. 
% JPO: is it exactly 10Hz or approximately? 
%Emily: It is approximately 10 Hz, but I just changed the text from "approximately 10 Hz" to "10 Hz." We can get into the details later in the paper.

There has been extensive work on using accelerometer data from wearable devices to determine a person's activity level. \citet{bai2014} proposed interpretable metrics to quantify how much a person moves, using accelerometer data. \citet{bai2012} proposed the so-called \emph{movelet method} to determine the activity that a person is performing based on raw data from a single, tri-axial accelerometer worn at the hip. \citet{he2014} applied the movelet method using data from three tri-axial accelerometers fixed at multiple points of the body, including the right hip, left wrist, and right wrist. \citet{xiao2016} considered activity classification for a person using another person's training data. % JPO: for which method?
\citet{urbanek2018} used the Fourier transform to detect sustained harmonic walking, defined as bouts of walking that are 10 seconds or more with low variability in step frequency, using body worn accelerometers. \citet{alhassan2008} analyzed ActiGraph accelerometer data collected during a field study, in which there was a missing data problem due to subjects sometimes not wearing the accelerometer. The authors proposed a method to estimate a subject's physical activity level using all of their available accelerometer data. There is also existing work on using both the accelerometer and gyroscope in wearable devices. With gyroscope and accelerometer data from wearable sensors, \citet{tahavori2017} applied machine learning methods, including random forest, support vector machine, LogitBoost, and Naive Bayes, to differentiate between six activities (tandem walk, stand to sit, sit to stand, stand, backwards walking, 3 meter walk), in healthy elderly and Parkinson's patients. 

The contributions of this paper include applying the movelet method proposed by \citet{bai2012} to smartphone data, while past work on this method focused on wearable device data. Smartphone data poses new challenges compared to wearable device data, such as having a lower frequency of data collection. For example, the ActiGraph wearable device can collect data at up to 100 Hz. The wearable sensors used in \citet{tahavori2017} collected data at 50 Hz. % JPO: need a sentence here to sum up smartphone data collection frequency to complete the comparison (wearables vs. phones)
%In addition, smartphone data can have planned missingness due to the sensors being turned off in order to preserve the battery of the phone or to reduce the volume of collected data. These gaps are planned since they are built into the data collection schedule that is chosen by the study investigator. 
Another contribution of this paper is that while previous work on the movelet method has focused solely on  accelerometer data, we demonstrate the potential of using both accelerometer and gyroscope data. The data used in this paper was collected in an experiment where four healthy participants performed various activities while wearing two smartphones and three wearable devices. The present paper focuses on accelerometer and gyroscope data from the two smartphones. %The collected raw data and all code discussed in this paper are available on GitHub. % JPO: github is usually not used for storing data and they don't really want to host data; we should identify another place. For example Harvard Dataverse is one option. Then there's another place whose name I cannot remember now, but they offer a DOI for the data, so it's time stamped and easily citable. I can ask Matt Kiang about this.
% Emily: Thanks, JP! I'll ask Matt about the other place.

The paper is organized as follows. Section \ref{data} presents some background on the study that we conducted. In Section \ref{methods}, we provide an overview of the movelet method developed by \citet{bai2012}. We apply the movelet method to our study data set and present the results in Section \ref{results}. Future work is discussed in Section \ref{discussion}.  

\section{Data}
\label{data}

\subsection{Study Procedure}

% JPO: need an opening here for why we collected our own data set vs. used somethng existing
%Currently, there are publicly available data sets for performing activity recognition using smartphone sensor data \citep{anguita2013, micucci2017, malekzadeh2018}. We conducted our own study because to our knowledge none of the available public data sets had all of our desired attributes: (1) collection of raw gyroscope and accelerometer data, (2) collection of data from different methods of carrying the phone (e.g., we use two smartphones in the front and back pants pockets, respectively, and reorient the front pocket phone in the four possible directions), (3) collection of wearable data in addition to smartphone data. 
Our study was approved by the Harvard T.H. Chan School of Public Health Institutional Review Board in February 2018 and the data collection took place in the summer of 2018. The eligibility requirements included being at least 18 years old, and able to walk, stand, and ascend and descend stairs without assistance from a person or device. Four healthy subjects, two females and two males, enrolled in the study. We collected data on age, weight, height, gender, dominant hand, and preferred method of carrying their personal phone (e.g., pocket, hand, bag). Table \ref{table:pt_characteristics} lists the participants' demographic characteristics. The participants ranged in age from 27 to 54.

Each participant completed an approximately one-hour study visit, during which she/he was outfitted with smartphones and wearable devices. The smartphones included an iPhone 5S in the front right pants pocket and an iPhone 7 in the back right pants pocket. Also, the participant wore ActiGraph GT9X Link wearable devices on the left wrist, right wrist, and right ankle, respectively. The smartphones collected gyroscope and accelerometer data continuously at a frequency of 10 Hz. % JPO: why approximately? Emily: I wrote "approximately 10 Hz" since the frequency isn't perfectly controlled at 10Hz; e.g., the spacing between subsequent data points is slightly less than 1 sec sometimes or slightly more than 1 sec at other times. I've taken out out the "approximately" since this is a minor detail. 
The ActiGraph devices collected gyroscope and accelerometer data continuously at 100 Hz. All study visits were videotaped, and we used the video footage to manually annotate each participant's accelerometer and gyroscope data sets with ground truth activity labels. We emphasize that this paper focuses on the smartphone data only, and a joint anlaysis of smartphone and wearable data will be presented elsewhere.

The participants performed a series of prescribed activities. We asked each participant to perform a series of activities to generate either training data for building a subject's dictionary (discussed in Section \ref{methods}) or test data for evaluating the accuracy of the dictionary-based classification model. During the collection of training data, the participant performed some routine activities for a short period of time. These included standing (for $\approx$ 10 seconds), walking on a flat surface ($\approx$ 15 meters), ascending a single flight of stairs, descending the flight of stairs, and performing two repetitions of chair stands (i.e., sitting down from standing, staying seated for 10 seconds, and standing up from sitting). During the collection of test data, the participant followed various routes on the Harvard Longwood campus that included walking, standing, sitting on benches, ascending stairs, and descending stairs. We asked the participant to complete one specific route four times, each time with the front pocket phone in a different orientation. We also collected data on the participant walking at different speeds, including ``normal'' (their normal speed), ``slow'' (slower than normal), and ``fast'' (faster than normal). To simulate a real life setting rather than a controlled lab environment, all data were collected in public places. In addition to training and test data, we collected data from an iPhone as the participant made a call and browsed the Internet using the phone. An outline of the complete study protocol is given in Table \ref{table:protocol}.

\section{Methods}
\label{methods}

The movelet method was proposed by \citet{bai2012} for activity classification using accelerometer data collected by wearable devices. This method only requires a small amount of training data and can be used to detect activities of the user's choice, even those that occur only for a brief moment, such as the transition from sitting to standing. In the method, the user first makes a list of common activities that occur in the subject's daily life, such as walking, ascending and descending stairs, standing, sitting, and running. The subject is then asked to perform each activity, and the resulting data is gathered. This data is referred to as training data, and only 2-3 seconds of training data are required per activity. The training data are then used to build a dictionary for the subject, whose entries are the different activities. Each entry consists of ``movelets,'' which are defined as 1-second windows of data. The movelets for a specific activity entry are obtained by sliding a 1-second window along the training data for the activity, starting with the left edge of the window at the first data point and sliding one data point at a time until the right edge of the window is at the last data point. The method assumes that the data is collected at a constant frequency. %JPO: how is the location of the last data point determined, i.e., how does one know when to stop sliding the window?

After the training data is collected, the subject goes about their daily life and new data without activity labels is collected. For the subject's new data, their dictionary is used to make classifications of the activity occurring at any given time point. First, each movelet in the new data set is compared to the dictionary movelets and the closest match is identified based on a distance metric, such as Euclidean distance. For example, if the closest match is to a dictionary movelet in the walk entry, the movelet in the new data set is classified to be walking. Second, for a given timepoint, a majority vote is taken among the neighbor movelets, including the movelet beginning at the time point and the movelets in the next second. The majority vote determines the predicted activity label at the time point. %JPO: does the standard method include uncertainty quantification? if not, we should think about how to incorporate it. so you have the observation for this point in time, call it x(t), and then you have it's distance to each movelet in the dictionary, call them d_1(x(t)) ... d_m(x(t)), if you have m movelets. If you order these distances such that d_1 \le d_2 ... \le d_m, your prediction is activity 1 since that's the one with the smallest distance (given how we've oredered them). But as d_1 and d_2 approach one another, you should be less and less certain that 1 is the right activity because your data looks more and more like activity 2. If it's an actual distance metric, then you're bound by 0 on the left, but what's the right bound or is there any? So the uncertainty should decrease as |d_1-d_2| increases, but that should be taken with respect to their sizes; in other words, a difference of 0.1 doesn't mean much is the distances are of the order of 10, but if they're of the order of 1, that's different. Something like 2*(d_2-d_1) / (d_1 + d_2) could be a starting point for thinking about uncertainty, but it would be nice to make a probabilistic statement. Anyway, we can thikn about this more. If this hasn't been done and we can do it, it would be a great addition to the paper and should't be hopefully too much extra work.

\section{Results}
\label{results}

\subsection{Training Data}

\begin{figure}[!p]
\centering\includegraphics[scale=0.7]{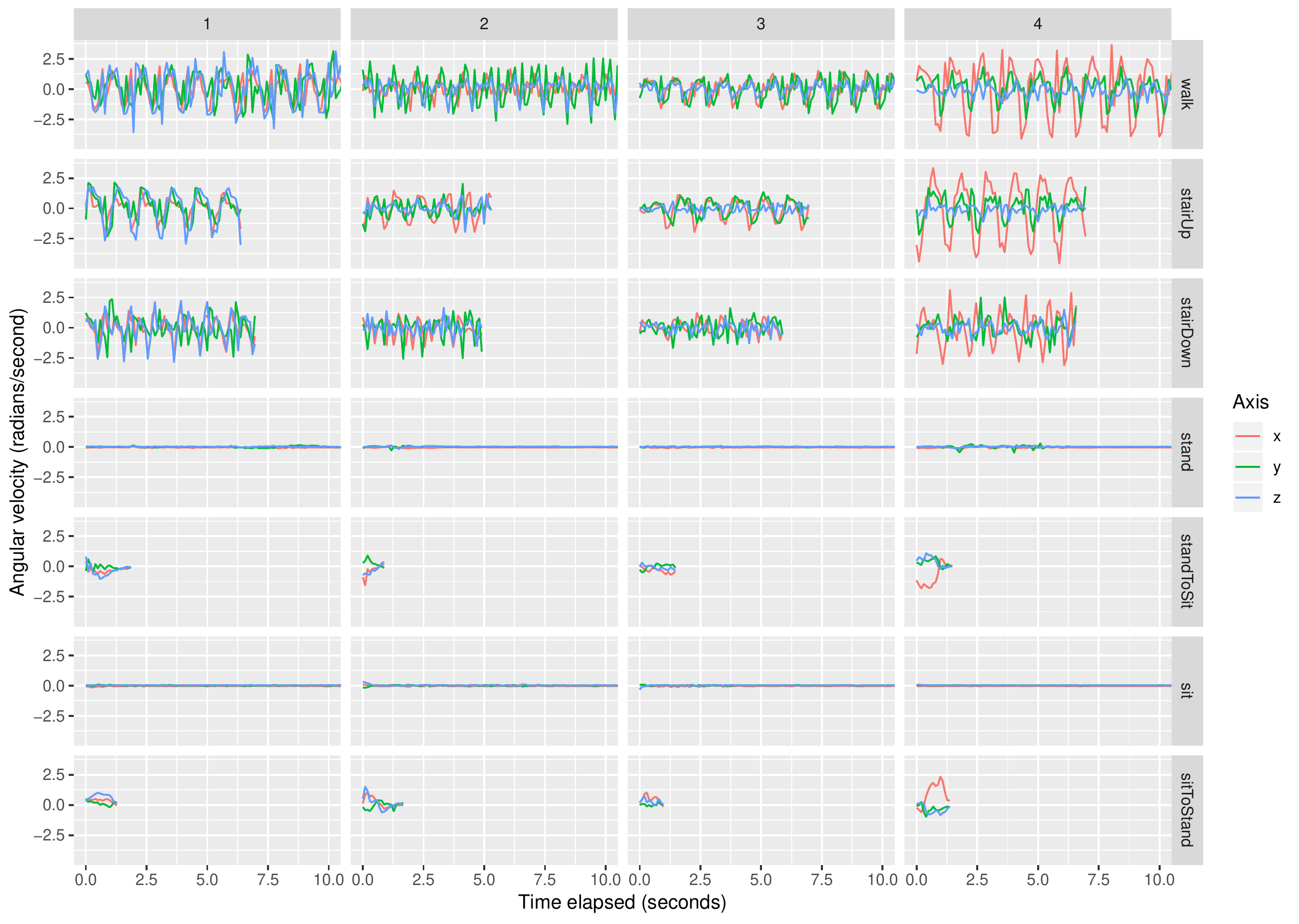}
\caption{Training data (tri-axial) from the front pocket smartphone gyroscope. \textit{The raw tri-axial data ($x$ = red, $y$ = green, $z$ = blue) from the front pocket gyroscope is shown. The columns indicate the participant ID number and the rows indicate the activity being performed. For the activities of ``standToSit,'' ``sit,'' and ``sitToStand,'' we plot the data from the first of the two chair-stands.}}
\label{training_tri_gyro_front}
\end{figure}

\begin{figure}[!p]
\centering\includegraphics[scale=0.7]{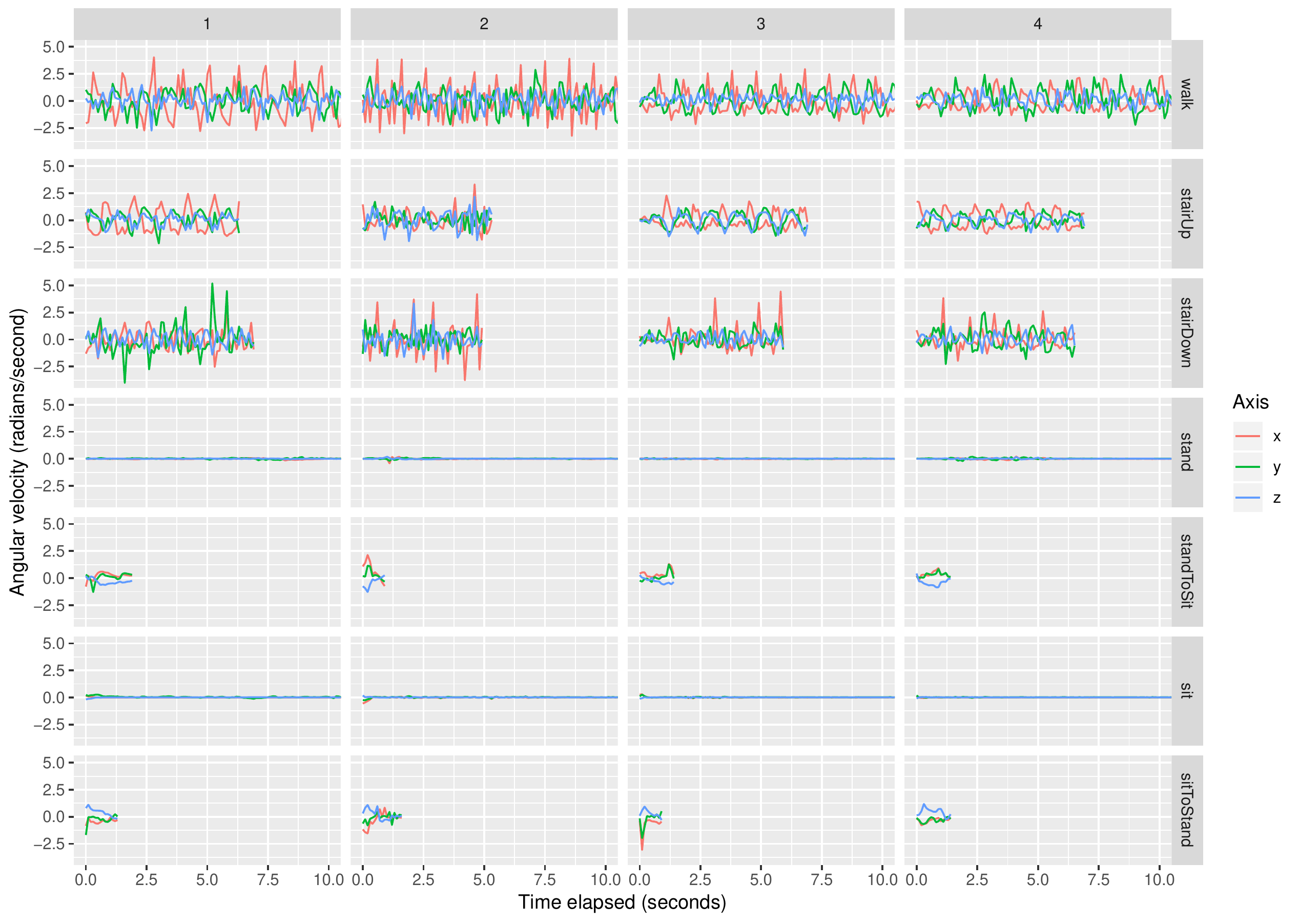}
\caption{Training data (tri-axial) from the back pocket smartphone gyroscope. \textit{The raw tri-axial data ($x$ = red, $y$ = green, $z$ = blue) from the back pocket gyroscope is shown. See Figure \ref{training_tri_gyro_front} for details.}}
\label{training_tri_gyro_back}
\end{figure}

\begin{figure}[!p]
\centering\includegraphics[scale=0.7]{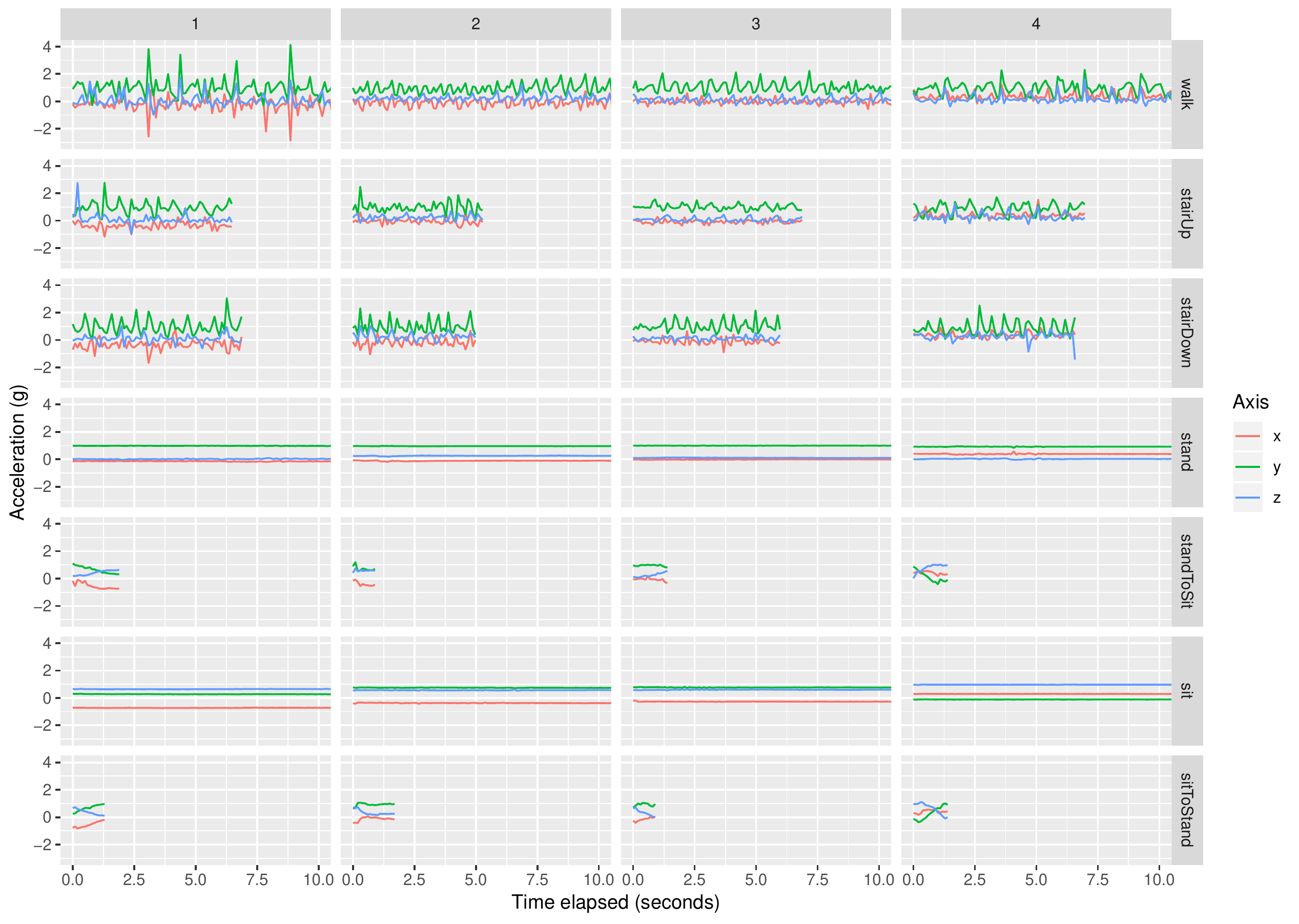}
\caption{Training data (tri-axial) from the front pocket smartphone accelerometer. \textit{The raw tri-axial data ($x$ = red, $y$ = green, $z$ = blue) from the front pocket accelerometer is shown. See Figure \ref{training_tri_gyro_front} for details.}}
\label{training_tri_acc_front}
\end{figure}

\begin{figure}[!p]
\centering\includegraphics[scale=0.7]{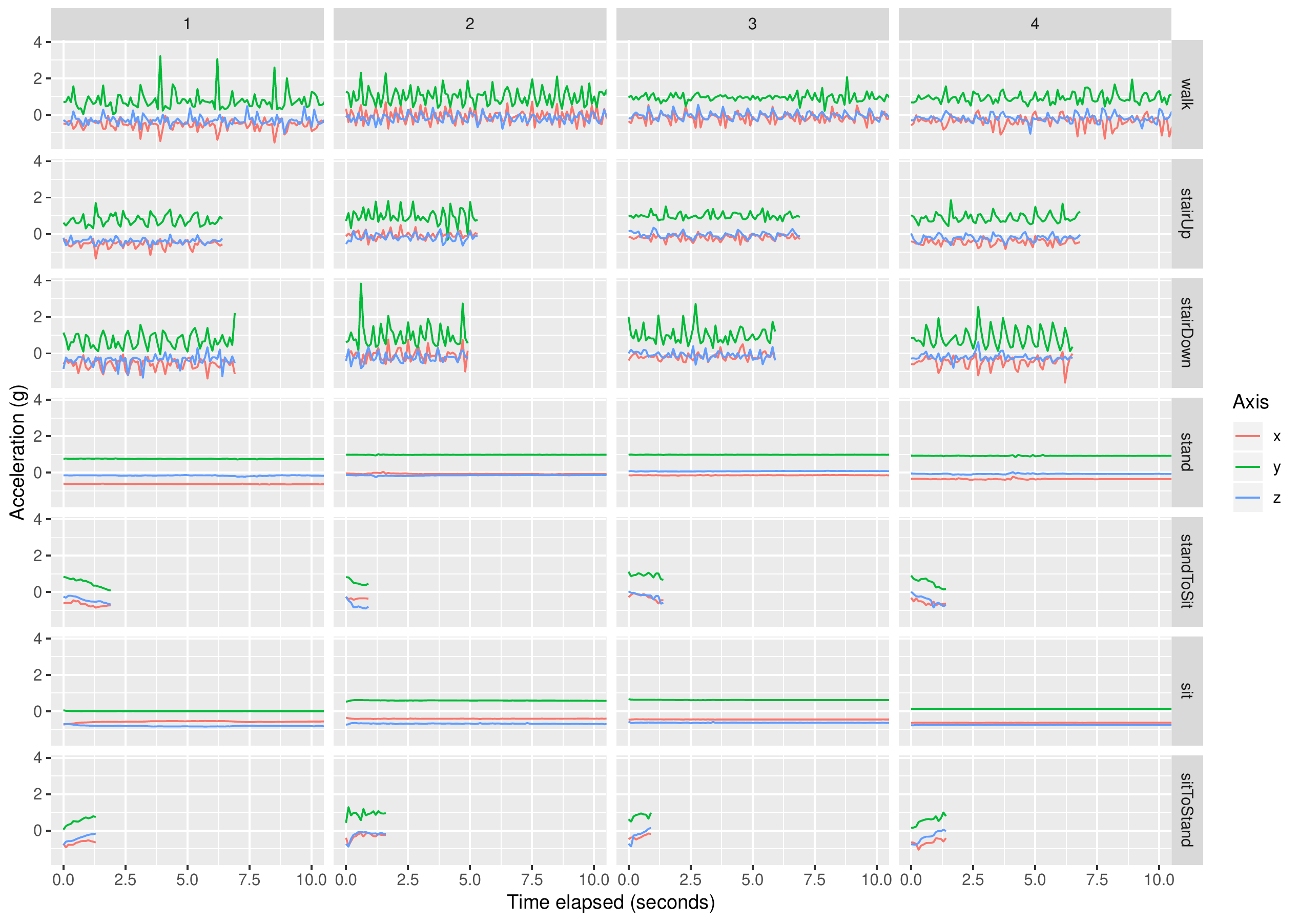}
\caption{Training data (tri-axial) from the back pocket smartphone accelerometer. \textit{The raw tri-axial data ($x$ = red, $y$ = green, $z$ = blue) from the back pocket accelerometer is shown. See Figure \ref{training_tri_gyro_front} for details.}}
\label{training_tri_acc_back}
\end{figure}

We first present the participants' training data. Training data was collected for standing, walking, ascending stairs, descending stairs, and chair-stands. Figure \ref{training_tri_gyro_front} presents the raw tri-axial (i.e., $x$, $y$, $z$) data from the front pocket smartphone gyroscope during training data collection. The four columns correspond to Participants 1-4, respectively. Each row corresponds to a specific activity. A complete chair stand was broken into three separate activities, including (i) stand-to-sit (ii) sit (iii) sit-to-stand. The plots of these three activities are taken from the first of the two chair stands that we asked the participant to perform. Figures \ref{training_tri_gyro_back}, \ref{training_tri_acc_front}, and \ref{training_tri_acc_back} have the same formatting as Figure \ref{training_tri_gyro_front}. They present the raw tri-axial data for the other sensors, including the back pocket smartphone gyroscope, front pocket smartphone accelerometer, and back pocket smartphone accelerometer, respectively. 

In general, we observe a fair bit of variability across the participants. For example, in Figure \ref{training_tri_gyro_front}, there are clear differences in the walking data across the subjects. For example, data for Participant 3 has a smaller amplitude than that for Participant 1, and for Participant 4 we see a large amplitude for the $x$ axis (shown in red) that is not present in the data for the other participants. There is also variability between the front and back pocket gyroscope data, as is evident for Participant 1. For all four participants, the gyroscope data from the back pocket appears more jittery than the gyroscope data from the front pocket during walking and stairs.  

For front and back pocket gyroscope data, the output during sitting and standing is roughly 0 radians per second because the phone is not rotating during either activity. Using the front pocket accelerometer, we can differentiate between sitting and standing because the phone is vertical during standing (so that gravity falls on the $y$ axis) while the phone is horizontal during sitting (so that gravity falls on the $z$ axis). For the back pocket accelerometer, the phone does not come to be horizontal when the participant is sitting, so here it is more difficult to differentiate between sitting and standing compared to the front pocket accelerometer. 

At any given timepoint $t$, let $\big(k_x(t), k_y(t), k_z(t)\big)$ denote the accelerometer or gyroscope measurement at time $t$. Then the magnitude is equal to $m(t) = \sqrt{k_x(t)^2 + k_y(t)^2 + k_z(t)^2}$. In the Supplementary Materials, we present plots of the magnitude data for the front gyroscope, back gyroscope, front accelerometer, and back accelerometer. % JPO: this paragraph stops short. Let's add a few sentences here to explain what we learn from these plots.

\subsection{Application of Movelet Method}

We applied the movelet method to the data we collected in this study. The method was applied to accelerometer and gyroscope data separately. For each participant, we built his/her dictionary using four seconds of training data per activity. If there were more than four seconds available, we used the middle four seconds. % JPO: am I reading this right, of the 1 hour of data that we collected, we're using just 4 seconds per activity for training.
%Emily: Yes, that's right!
The list of activities include those along the right hand margin of Figure \ref{training_tri_gyro_front}. To do classifications on test data from the front gyroscope, we used the dictionary corresponding to the front gyroscope. The handling for the back gyroscope, front accelerometer, and back accelerometer was analogous. When comparing movelets in the test data to movelets in the dictionary, we used Euclidean distance as the distance metric. As a sensitivity analysis, we show the results if the raw tri-axial data is used, as well as the results if only the magnitude data is used. 

Tables \ref{table:results_frontgyro}, \ref{table:results_backgyro}, \ref{table:results_frontacc}, and \ref{table:results_backacc} compare the predicted activity label to the true activity label for each participant. The separate tables show results for the front gyroscope, back gyroscope, front accelerometer, and back accelerometer, respectively. The results consider all portions of the test data collection, except for when the participant walked at different speeds (discussed at the end of this subsection) and when the participant performed the same course multiple times with the front pocket phone in a different orientation each time (discussed in Section \ref{phone_orientation}). The table rows give the predicted activity labels, while the columns give the true activity labels. For any given row and column, there are two values that each represent the proportion of times that the activity given by the column is predicted to be the activity given by the row. The value on the left is for using tri-axial data, and the the value on the right is for using magnitude.

Using the movelet method, the classification results for walking, ascending stairs, and descending stairs were strong for the gyroscope (Tables \ref{table:results_frontgyro} and \ref{table:results_backgyro}). For the front pocket gyroscope (Tables \ref{table:results_frontgyro}), the classification accuracies when tri-axial data was used ranged from 0.72 to 0.95 for walking, 0.76 to 1 for ascending stairs, and 0.7 to 0.88 for descending stairs. The classification accuracies when magnitude data was used ranged from 0.65 to 0.91 for walking, 0.73 to 0.98 for ascending stairs, and 0.15* to 0.87 for descending stairs. (*For Participant 4, using tri-axial data yielded a better accuracy of 0.81 compared to 0.15 using magnitude data.) For the back pocket gyroscope (Table \ref{table:results_backgyro}), the classification accuracies when tri-axial data was used ranged from 0.82 to 0.92 for walking, 0.66 to 1 for ascending stairs, and 0.46 to 0.85 for descending stairs. The classification accuracies when magnitude data was used ranged from 0.81 to 0.94 for walking, 0.78 to 0.99 for ascending stairs, and 0.32 to 0.67 for descending stairs. As a comparison, the null rate, which is based on random guessing, is 1/number of training activities = 0.14. % JPO: somewhere, perhaps at the beginning of this paragraph, we should state what the null rate is, i.e. 1 / number of activities. That will give us a baseline number based on guessing alone.

The gyroscope data outperformed its accelerometer counterpart in correctly predicting walking, ascending stairs, and descending stairs. For the front pocket accelerometer (Table \ref{table:results_frontacc}), the classification accuracies when tri-axial data was used ranged from 0.57 to 0.86 for walking, 0.28 to 0.86 for ascending stairs, and 0.36 to 0.56 for descending stairs. For the accelerometer, using magnitude data improved the results compared to using tri-axial data. The classification accuracies when magnitude data was used ranged from 0.62 to 0.92 for walking, 0.38 to 0.97 for ascending stairs, and 0.38 to 0.78 for descending stairs. For the back pocket accelerometer (Table \ref{table:results_backacc}), the classification accuracies when tri-axial data was used ranged from 0.36 to 0.71 for walking, 0 to 0.74 for ascending stairs, 0.06 to 0.69 for descending stairs. The classification accuracies when magnitude data was used ranged from 0.65 to 0.93 for walking, 0.41 to 0.95 for ascending stairs, and 0.46 to 0.65 for descending stairs.

We asked the participant to walk at various speeds, including ``normal,'' ``fast,'' and ``slow.'' Tables \ref{table:walking_speeds_TRI} and \ref{table:walking_speeds_MAG} present the distribution of the predicted activity label for each data type and each participant, under the three walking speeds. Table \ref{table:walking_speeds_TRI} shows the results when raw tri-axial data is used. At the normal pace, the front and back gyroscopes both performed well. For participants 2 and 4, the gyroscopes predicted walking 100\% of the time. Slow walking was sometimes mistaken for stairs, in particular ascending stairs. The gyroscope data was able to recognize fast walking as walking, especially the front gyroscope for participants 2 and 4 and the back gyroscope for participants 1 and 3. In general, the gyroscopes outperformed their accelerometer counterparts at recognizing walking at different speeds. Table \ref{table:walking_speeds_MAG} shows the results when only magnitude data is used. Using magnitude helped the back accelerometer predict walking. For the gyroscope, the classification accuracy is worse for slow walking but better for fast walking compared to using tri-axial data.

\subsection{Phone Orientation}
\label{phone_orientation}

A phone can be placed in one of four possible orientations inside the pants pocket, given by whether the phone screen is facing the leg or not, and whether or not the phone is upside down. %Table \ref{table:transform} shows how to transform raw tri-axial data to a standard frame of reference (i.e., that the phone is upside down and with the face against the leg). % JPO: need a bit more text here, otherwise seems like the paragraph stopped early
During the data collection, we asked the participant to repeat four times a course that included standing, walking, and stairs. Each time, the phone in the front pocket was re-oriented, so that data from each possible orientation was observed. For the test data collected during this segment, we implemented the movelet method under two scenarios: (1) using the raw tri-axial data without any adjustment, even though the training data was collected under a single orientation, (2) using magnitude data. Tables \ref{table:vary_orientation_frontgyro} and \ref{table:vary_orientation_frontacc} show the distribution of the predicted activity label for each activity during this segment of the test data collection. The separate tables are for the front pocket gyroscope and the front pocket accelerometer, respectively. We focus on the front pocket phone because the back pocket phone was not re-oriented. The table rows show the predicted activity labels, and the columns give the true activity labels. For any given row and column, the two values represent the proportion of times that the activity given by the column is classified to be the activity given by the row. The first value is the proportion if raw tri-axial data is used; the second value is the proportion if only magnitude data is used. 

For the front pocket gyroscope (Table \ref{table:vary_orientation_frontgyro}), the classification accuracy for walking ranges from 0.74 to 0.90 when magnitude data is used, and is not as high when tri-axial data is used (0.38 to 0.87). The classification accuracy for ascending stairs is high in either case, 0.70 to 0.94 for tri-axial data and 0.61 to 1 for magnitude data. For descending stairs, the classification accuracy of using tri-axial data versus using magnitude data depends on the participant, e.g., 0.43 versus 0.80 for Participant 1 whereas 0.91 versus 0.51 for Participant 3. Standing is sometimes confused for sitting, which is expected since the phone is not rotating in either case, so it is difficult for the gyroscope to differentiate between these two stationary activities. % In this paragraph and the paragraph below, it is not clear that these numbers contrast different phone orientations. I think we can address this by adding "across different orientations" or similar in a couple of places.

The front pocket accelerometer (Table \ref{table:vary_orientation_frontacc}) can differentiate standing from sitting. However, the relative performance of using tri-axial compared to magnitude data differs by participant. For example, the classification accuracy during standing is 0 (tri-axial) versus 0.92 (magnitude) for Participant 4, while it is 0.94 (tri-axial) versus 0.55 (magnitude) for Participant 1.  For walking, using magnitude data works better on average across the participants, yielding classification accuracies between 0.52 to 0.95. Using magnitude data also helps for recognizing climbing stairs, yielding classification accuracies between 0.34 to 0.98. The same holds for descending stairs, where the classification accuracies when magnitude was used ranged from 0.53 to 0.86.

\section{Discussion}
\label{discussion}

We applied the movelet method to smartphone gyroscope and accelerometer data from our study of healthy volunteers. In the study, data was collected in public places rather than a tightly controlled lab environment, to mimic data collection in the wild. Ground truth activity labels based on video footage were used to validate the activity predictions. Using the gyroscope data, the method generally predicted walking, ascending stairs, and descending stairs with a high sensitivity. Also, we could classify fast walking correctly as walking with high sensitivity, even though the training data collection did not include fast walking. The prediction results for the gyroscope were generally better than those for the accelerometer. However, the accelerometer was more reliable for differentiating between stationary activities, such as standing versus sitting. For the accelerometer, the classification accuracies were generally improved by applying the movelet method to magnitude data compared to tri-axial data.

An advantage of the movelet method is that it requires only a small amount of training data to build a person's dictionary. In our analyses, we used just four seconds of training data per activity for each participant's dictionary. However, in large studies it may be challenging to collect labeled training data on every participant. One option is to match each participant without labeled training data to another person for whom labeled training data is available, and this matching can be based on variables such as age, height, weight, gender, and preferred phone carrying position. An area of future research is evaluate the accuracy of this approach. Also, one way to streamline the collection of training data is to incorporate the data collection into routine tests that are already conducted during clinic visits, such as the six-minute walk test. % JPO: so this test is almost never conducted, although it's very relevant for cardiovascular health; let's thikn about other possible examples

In this paper, we focused on the case that the phone is in a pants pocket. Another challenge is to consider other possible placements of the phone.  The study data showed that the placement of the phone in the front pocket compared to the back pocket affected the data and subsequent activity classification. The data also will look different if the phone is in the hand, a backpack, or a purse. A potential area of future research is to extend the movelet method to handle unknown and changing placements of the phone.% by augmenting the dictionary with entries for each potential method of carrying the phone, e.g., by having multiple entries for walking like ``walking-pocket,'' ``walking-backpack,'' and ``walking-hand.''  However, this will make the training data collection longer. It also would increase the required computation time because there will be more movelets in the participant's dictionary. A remedy for the latter is to consider parallelizing the code, such as by doing classifications for different timepoints on different nodes of a cluster. 

\section{Supplementary Material}
\label{suppMat}

The Supplementary Materials are available upon request. Please contact Emily Huang at ehuang@hsph.harvard.edu.

\bibliographystyle{biom}
\bibliography{refs}

%\begin{figure}[p!]
%\centering
%\begin{subfigure}[b]{0.7\textwidth}
%\centering
%\includegraphics[width = 0.9\textwidth]{acc_schematic.png}
%\caption{Accelerometer}
%\end{subfigure}
%\begin{subfigure}[b]{0.7\textwidth}
%\centering
%\includegraphics[width = 0.9\textwidth]{gyro_schematic.png}
%\caption{Gyroscope}
%\end{subfigure}
%\caption{Schematics of the Accelerometer and Gyroscope Sensors (from Apple %\label{sensor_schematics}
%\end{figure}

\clearpage

\begin{table}[!p]
\caption{Outline of Study Protocol.
\label{table:protocol}}
    \centering
    \begin{tabular}{cc}
\hline
 & Description  \\
\hline 
 Using the Phone &
    \small
 \parbox{.75\textwidth}{\begin{enumerate}
    \item The participant puts on three Actigraph wearable devices, tightening the straps according to their preference. There is a wearable on the right ankle and one on each wrist.
    \item The study investigator puts a smartphone on a table. The participant picks up the phone from the table, finds the study investigator's name under ``Contacts,'' calls the study investigator holding the phone to their ear, hangs up after reaching the voicemail greeting, and sets the phone back on the table.
    \item Starting with the phone on the table, the participant picks up the phone from the table, presses the icon for the Safari browser, enters the word ``statistics'' into the search bar, selects the Wikipedia page that appears as the first search result, scrolls through the Wikipedia entry, and sets the phone back on the table.
    \item The participant puts one smartphone in their front pants pocket and the other smartphone in their back pants pocket.
\end{enumerate}} \\
\hline 
Training Data &
\small\parbox{.75\textwidth}{\begin{enumerate}
    \item Standing for $\approx$ 10 seconds
    \item Walking on a flat surface for $\approx$ 15 meters
    \item Ascending one flight of stairs
    \item Descending one flight of stairs 
    \item Two chair-stands (i.e., sitting down from standing, staying seated for $\approx$ 10 seconds, and standing up from sitting)
\end{enumerate}} \\
\hline
Test Data & \small\parbox{.75\textwidth}{\begin{enumerate}
    \item Going around the Harvard Medical School Quadrangle (which consists of walking, ascending stairs, descending stairs, and standing).
    \item Walking to a bench and sitting down, then walking to another bench and sitting down.
    \item Walking at different speeds (normal, fast, then slow).
    \item Repeating (1) four times, each time with the front pocket phone in a different orientation.
    \item Walking from the Harvard Medical School Quadrangle into the Kresge building of the Harvard School of Public Health (which consists of walking, ascending stairs, descending stairs, standing, and going through a revolving door to enter the building).
    \item Descending and ascending a flight of stairs in the Kresge building. This staircase was steeper than other staircases used in the data collection.
\end{enumerate}} \\
\hline
\end{tabular}
\end{table}

\clearpage

\begin{table}[!p]
\caption{Participant Characteristics.
\label{table:pt_characteristics}}
\centering
\small\addtolength{\tabcolsep}{-2pt}
\resizebox{\columnwidth}{!}{%
\begin{tabular}{ccccccc}
\hline
Participant & Gender & Age & Height & Weight & Dominant Hand & Phone Carrying Preference\\
\hline
1 & male & 41  & 1.83 m (6 ft 0 in.) & 86 kg (190 lb) & right & pants pocket \tabularnewline
2 & female & 54  & 1.75 m (5 ft 9 in.)  & 79 kg (175 lb) & right & {purse, hand, pants pocket} \tabularnewline
3 & female & 27 & 1.60 m (5 ft 3 in.) & 57 kg (125 lb) & right & {purse, hand, pants pocket} \tabularnewline
4 & male & 28 &  1.88 m (6 ft 2 in.) & 84 kg (185 lb) & right & {pants pocket}\\
\hline
\end{tabular}
}
\textit{Abbreviation Key: m = meters, ft = feet, in = inches, kg = kilograms, lb = pounds}
\end{table}

\clearpage

% JPO: overall accuracy for each data type (magnitude vs. 3-ax.) as an average of the diagonal elements; report both averages for each person in the caption

% JPO: in the tables below, does the formatting require us to use heading case for the first sentence in the caption? we don't do that for figure captions, so for consistency, unless required by the journal we're considering, I'd also make the first sentences to have regular sentence case.

% JPO: What is revolveDoor, the last column in the table? Note that that's noc included as a row, so that's a bit confusing.
% Emily: In the caption, I have added a definition for revolveDoor and clarified why it's not included as a row.

% JPO: on these lines, e.g., Participant 1 (0.76, 0.78; 0.82, 0.89), what are these numbers? I'm not sure we're explained them
\begin{table}
\caption{True Activity Label versus Predicted Activity Label for Front Pocket Smartphone Gyroscope. \textit{These results incorporate all segments of the test data collection, except for walking at different speeds and flipping the phone around. Subtables (a)-(d) correspond to Participants 1-4, respectively. In each subtable, the column headings show the true activity label and the row headings show the predicted activity label. There are two subcolumns under each column heading, the first (on the left) for classifications based on tri-axial data and the second (on the right) for classifications based on magnitude data. Each subcolumn gives the distribution of predicted activity labels, when the true activity label is that given by the column heading. Thus, each subcolumn sums to 1. The shaded diagonal highlights the proportion of activity label classifications that match the true activity label. The closer these diagonals are to 1 (the perfect score), the better the accuracy of the classifier. The average of the diagonal elements for each data type (tri-axial vs. magnitude) is 0.76 vs. 0.78 for Participant 1, 0.63 vs. 0.39 for Participant 2, 0.65 vs. 0.50 for Participant 3, and 0.61 vs. 0.51 for Participant 4. The  ``revolveDoor'' activity refers to going through a revolving door and is not a row heading since it was not part of the training data collection.} \label{table:results_frontgyro}}
\centering
\scalebox{0.8}{
\small\addtolength{\tabcolsep}{-2pt}
\begin{tabular}{rllllllllllllllll}
  \hline
  & \multicolumn{16}{c}{\bf{(a) Participant 1 }} \\%(0.76, 0.78; 0.82, 0.89)
    \hline
  \tabularnewline
  & \multicolumn{2}{c}{stand} & \multicolumn{2}{c}{walk} & \multicolumn{2}{c}{stairUp} & \multicolumn{2}{c}{stairDown} & \multicolumn{2}{c}{standToSit} & \multicolumn{2}{c}{sit} & \multicolumn{2}{c}{sitToStand} & \multicolumn{2}{c}{revolveDoor} \\
  stand & \cellcolor[gray]{0.9}{0.31} & \cellcolor[gray]{0.9}{0.45} & 0.00 & 0.00 & 0.00 & 0.00 & 0.00 & 0.00 & 0.05 & 0.10 & 0.11 & 0.16 & 0.00 & 0.00 & 0.00 & 0.00 \\ 
  walk & 0.01 & 0.00 & \cellcolor[gray]{0.9}{0.72} & \cellcolor[gray]{0.9}{0.84} & 0.00 & 0.02 & 0.07 & 0.04 & 0.00 & 0.00 & 0.00 & 0.00 & 0.00 & 0.00 & 0.00 & 0.00 \\ 
  stairUp & 0.06 & 0.05 & 0.20 & 0.14 & \cellcolor[gray]{0.9}{1.00} & \cellcolor[gray]{0.9}{0.98} & 0.18 & 0.07 & 0.00 & 0.00 & 0.00 & 0.00 & 0.00 & 0.00 & 0.80 & 0.27 \\ 
  stairDown & 0.04 & 0.00 & 0.07 & 0.01 & 0.00 & 0.00 & \cellcolor[gray]{0.9}{0.76} & \cellcolor[gray]{0.9}{0.87} & 0.00 & 0.00 & 0.00 & 0.00 & 0.05 & 0.00 & 0.20 & 0.04 \\ 
  standToSit & 0.00 & 0.04 & 0.01 & 0.00 & 0.00 & 0.00 & 0.00 & 0.00 & \cellcolor[gray]{0.9}{0.75} & \cellcolor[gray]{0.9}{0.75} & 0.00 & 0.02 & 0.00 & 0.24 & 0.00 & 0.20 \\ 
  sit & 0.57 & 0.42 & 0.00 & 0.00 & 0.00 & 0.00 & 0.00 & 0.00 & 0.20 & 0.00 & \cellcolor[gray]{0.9}{0.84} & \cellcolor[gray]{0.9}{0.80} & 0.00 & 0.00 & 0.00 & 0.00 \\ 
  sitToStand & 0.00 & 0.03 & 0.00 & 0.01 & 0.00 & 0.00 & 0.00 & 0.02 & 0.00 & 0.15 & 0.05 & 0.03 & \cellcolor[gray]{0.9}{0.95} & \cellcolor[gray]{0.9}{0.76} & 0.00 & 0.49 \\ 
  \tabularnewline
  \hline
  
    & \multicolumn{16}{c}{\bf{(b) Participant 2 }} \\%(0.63, 0.39; 0.85, 0.64)
    \hline
  \tabularnewline
  & \multicolumn{2}{c}{stand} & \multicolumn{2}{c}{walk} & \multicolumn{2}{c}{stairUp} & \multicolumn{2}{c}{stairDown} & \multicolumn{2}{c}{standToSit} & \multicolumn{2}{c}{sit} & \multicolumn{2}{c}{sitToStand} & \multicolumn{2}{c}{revolveDoor} \\
  stand & \cellcolor[gray]{0.9}{0.20} & \cellcolor[gray]{0.9}{0.00} & 0.00 & 0.00 & 0.00 & 0.00 & 0.00 & 0.00 & 0.00 & 0.00 & 0.16 & 0.03 & 0.00 & 0.00 & 0.00 & 0.00 \\ 
  walk & 0.01 & 0.00 & \cellcolor[gray]{0.9}{0.93} & \cellcolor[gray]{0.9}{0.65} & 0.06 & 0.05 & 0.25 & 0.19 & 0.00 & 0.00 & 0.00 & 0.00 & 0.48 & 0.52 & 0.16 & 0.00 \\ 
  stairUp & 0.00 & 0.03 & 0.02 & 0.30 & \cellcolor[gray]{0.9}{0.91} & \cellcolor[gray]{0.9}{0.83} & 0.05 & 0.36 & 0.00 & 0.60 & 0.00 & 0.01 & 0.05 & 0.48 & 0.42 & 1.00 \\ 
  stairDown & 0.00 & 0.00 & 0.01 & 0.05 & 0.00 & 0.09 & \cellcolor[gray]{0.9}{0.70} & \cellcolor[gray]{0.9}{0.43} & 0.00 & 0.00 & 0.00 & 0.03 & 0.00 & 0.00 & 0.00 & 0.00 \\ 
  standToSit & 0.00 & 0.03 & 0.02 & 0.00 & 0.00 & 0.02 & 0.00 & 0.02 & \cellcolor[gray]{0.9}{0.45} & \cellcolor[gray]{0.9}{0.00} & 0.00 & 0.06 & 0.00 & 0.00 & 0.00 & 0.00 \\ 
  sit & 0.73 & 0.78 & 0.00 & 0.00 & 0.00 & 0.00 & 0.00 & 0.00 & 0.05 & 0.00 & \cellcolor[gray]{0.9}{0.74} & \cellcolor[gray]{0.9}{0.83} & 0.00 & 0.00 & 0.00 & 0.00 \\ 
  sitToStand & 0.07 & 0.16 & 0.01 & 0.00 & 0.03 & 0.00 & 0.00 & 0.00 & 0.50 & 0.40 & 0.10 & 0.03 & \cellcolor[gray]{0.9}{0.48} & \cellcolor[gray]{0.9}{0.00} & 0.42 & 0.00 \\ 
  \tabularnewline
  \hline
  
    & \multicolumn{16}{c}{\bf{(c) Participant 3 }} \\%(0.65, 0.50; 0.87, 0.74)
    \hline
  \tabularnewline
  & \multicolumn{2}{c}{stand} & \multicolumn{2}{c}{walk} & \multicolumn{2}{c}{stairUp} & \multicolumn{2}{c}{stairDown} & \multicolumn{2}{c}{standToSit} & \multicolumn{2}{c}{sit} & \multicolumn{2}{c}{sitToStand} & \multicolumn{2}{c}{revolveDoor} \\
  stand & \cellcolor[gray]{0.9}{0.25} & \cellcolor[gray]{0.9}{0.60} & 0.00 & 0.00 & 0.00 & 0.00 & 0.00 & 0.00 & 0.00 & 0.00 & 0.23 & 0.26 & 0.00 & 0.00 & 0.00 & 0.00 \\ 
  walk & 0.00 & 0.00 & \cellcolor[gray]{0.9}{0.73} & \cellcolor[gray]{0.9}{0.84} & 0.00 & 0.17 & 0.10 & 0.16 & 0.00 & 0.25 & 0.01 & 0.02 & 0.15 & 0.95 & 0.00 & 0.00 \\ 
  stairUp & 0.00 & 0.03 & 0.22 & 0.04 & \cellcolor[gray]{0.9}{1.00} & \cellcolor[gray]{0.9}{0.83} & 0.02 & 0.25 & 0.00 & 0.40 & 0.03 & 0.05 & 0.60 & 0.05 & 0.90 & 0.95 \\ 
  stairDown & 0.10 & 0.00 & 0.03 & 0.10 & 0.00 & 0.00 & \cellcolor[gray]{0.9}{0.88} & \cellcolor[gray]{0.9}{0.54} & 0.00 & 0.25 & 0.00 & 0.02 & 0.25 & 0.00 & 0.10 & 0.05 \\ 
  standToSit & 0.02 & 0.24 & 0.00 & 0.01 & 0.00 & 0.00 & 0.01 & 0.05 & \cellcolor[gray]{0.9}{1.00} & \cellcolor[gray]{0.9}{0.10} & 0.02 & 0.04 & 0.00 & 0.00 & 0.00 & 0.00 \\ 
  sit & 0.62 & 0.14 & 0.01 & 0.00 & 0.00 & 0.00 & 0.00 & 0.00 & 0.00 & 0.00 & \cellcolor[gray]{0.9}{0.67} & \cellcolor[gray]{0.9}{0.61} & 0.00 & 0.00 & 0.00 & 0.00 \\ 
  sitToStand & 0.00 & 0.00 & 0.00 & 0.00 & 0.00 & 0.00 & 0.00 & 0.00 & 0.00 & 0.00 & 0.03 & 0.00 & \cellcolor[gray]{0.9}{0.00} & \cellcolor[gray]{0.9}{0.00} & 0.00 & 0.00 \\ 
  \tabularnewline
  \hline
  
    & \multicolumn{16}{c}{\bf{(d) Participant 4 }} \\%(0.61, 0.51; 0.84, 0.60)
    \hline
  \tabularnewline
  & \multicolumn{2}{c}{stand} & \multicolumn{2}{c}{walk} & \multicolumn{2}{c}{stairUp} & \multicolumn{2}{c}{stairDown} & \multicolumn{2}{c}{standToSit} & \multicolumn{2}{c}{sit} & \multicolumn{2}{c}{sitToStand} & \multicolumn{2}{c}{revolveDoor} \\
  stand & \cellcolor[gray]{0.9}{0.51} & \cellcolor[gray]{0.9}{0.62} & 0.01 & 0.00 & 0.00 & 0.00 & 0.00 & 0.00 & 0.30 & 0.30 & 0.40 & 0.41 & 0.00 & 0.00 & 0.00 & 0.00 \\ 
  walk & 0.00 & 0.00 & \cellcolor[gray]{0.9}{0.95} & \cellcolor[gray]{0.9}{0.91} & 0.23 & 0.23 & 0.17 & 0.62 & 0.00 & 0.00 & 0.00 & 0.00 & 0.00 & 0.52 & 0.23 & 0.10 \\ 
  stairUp & 0.00 & 0.00 & 0.01 & 0.05 & \cellcolor[gray]{0.9}{0.76} & \cellcolor[gray]{0.9}{0.73} & 0.02 & 0.21 & 0.00 & 0.00 & 0.00 & 0.00 & 0.00 & 0.10 & 0.00 & 0.00 \\ 
  stairDown & 0.24 & 0.05 & 0.02 & 0.02 & 0.01 & 0.04 & \cellcolor[gray]{0.9}{0.81} & \cellcolor[gray]{0.9}{0.15} & 0.00 & 0.00 & 0.02 & 0.02 & 0.95 & 0.33 & 0.77 & 0.90 \\ 
  standToSit & 0.00 & 0.05 & 0.00 & 0.01 & 0.00 & 0.00 & 0.00 & 0.00 & \cellcolor[gray]{0.9}{0.70} & \cellcolor[gray]{0.9}{0.70} & 0.00 & 0.04 & 0.00 & 0.05 & 0.00 & 0.00 \\ 
  sit & 0.25 & 0.15 & 0.00 & 0.00 & 0.00 & 0.00 & 0.00 & 0.00 & 0.00 & 0.00 & \cellcolor[gray]{0.9}{0.49} & \cellcolor[gray]{0.9}{0.48} & 0.00 & 0.00 & 0.00 & 0.00 \\ 
  sitToStand & 0.00 & 0.12 & 0.00 & 0.00 & 0.00 & 0.00 & 0.00 & 0.02 & 0.00 & 0.00 & 0.08 & 0.04 & \cellcolor[gray]{0.9}{0.05} & \cellcolor[gray]{0.9}{0.00} & 0.00 & 0.00 \\ 
  \hline
  \end{tabular}
  }
  \end{table}

  \begin{table}
\caption{True Activity Label versus Predicted Activity Label for Back Pocket Smartphone Gyroscope. \textit{See Table \ref{table:results_frontgyro} for details. The average of the diagonal elements for each data type (tri-axial vs. magnitude) is 0.86 vs. 0.56 for Participant 1, 0.45 vs. 0.44 for Participant 2, 0.71 vs. 0.51 for Participant 3, and 0.58 vs. 0.44 for Participant 4.}
\label{table:results_backgyro}  }
\centering
\resizebox{\columnwidth}{!}{%
\small\addtolength{\tabcolsep}{-2pt}
\begin{tabular}{rllllllllllllllll}
  \hline
  & \multicolumn{16}{c}{\bf{(a) Participant 1 }} \\%(0.86, 0.56; 0.90, 0.82)
    \hline
  \tabularnewline
  & \multicolumn{2}{c}{stand} & \multicolumn{2}{c}{walk} & \multicolumn{2}{c}{stairUp} & \multicolumn{2}{c}{stairDown} & \multicolumn{2}{c}{standToSit} & \multicolumn{2}{c}{sit} & \multicolumn{2}{c}{sitToStand} & \multicolumn{2}{c}{revolveDoor} \\
  stand & \cellcolor[gray]{0.9}{0.87} & \cellcolor[gray]{0.9}{0.38} & 0.00 & 0.00 & 0.00 & 0.00 & 0.03 & 0.00 & 0.15 & 0.00 & 0.13 & 0.07 & 0.20 & 0.00 & 0.00 & 0.00 \\ 
  walk & 0.00 & 0.00 & \cellcolor[gray]{0.9}{0.85} & \cellcolor[gray]{0.9}{0.82} & 0.00 & 0.00 & 0.08 & 0.10 & 0.00 & 0.00 & 0.00 & 0.00 & 0.00 & 0.00 & 0.00 & 0.14 \\ 
  stairUp & 0.09 & 0.07 & 0.13 & 0.07 & \cellcolor[gray]{0.9}{1.00} & \cellcolor[gray]{0.9}{0.99} & 0.04 & 0.21 & 0.00 & 0.00 & 0.00 & 0.00 & 0.00 & 0.00 & 0.26 & 0.20 \\ 
  stairDown & 0.00 & 0.04 & 0.00 & 0.11 & 0.00 & 0.01 & \cellcolor[gray]{0.9}{0.85} & \cellcolor[gray]{0.9}{0.67} & 0.00 & 0.00 & 0.00 & 0.00 & 0.00 & 0.00 & 0.20 & 0.10 \\ 
  standToSit & 0.04 & 0.04 & 0.01 & 0.01 & 0.00 & 0.00 & 0.01 & 0.03 & \cellcolor[gray]{0.9}{0.85} & \cellcolor[gray]{0.9}{0.20} & 0.01 & 0.08 & 0.00 & 1.00 & 0.30 & 0.56 \\ 
  sit & 0.00 & 0.48 & 0.00 & 0.00 & 0.00 & 0.00 & 0.00 & 0.00 & 0.00 & 0.05 & \cellcolor[gray]{0.9}{0.82} & \cellcolor[gray]{0.9}{0.84} & 0.00 & 0.00 & 0.00 & 0.00 \\ 
  sitToStand & 0.00 & 0.00 & 0.01 & 0.00 & 0.00 & 0.00 & 0.00 & 0.00 & 0.00 & 0.75 & 0.04 & 0.01 & \cellcolor[gray]{0.9}{0.80} & \cellcolor[gray]{0.9}{0.00} & 0.24 & 0.00 \\
  \tabularnewline
  \hline
  
    & \multicolumn{16}{c}{\bf{(b) Participant 2 }} \\%(0.45, 0.44; 0.71, 0.72)
    \hline
  \tabularnewline
  & \multicolumn{2}{c}{stand} & \multicolumn{2}{c}{walk} & \multicolumn{2}{c}{stairUp} & \multicolumn{2}{c}{stairDown} & \multicolumn{2}{c}{standToSit} & \multicolumn{2}{c}{sit} & \multicolumn{2}{c}{sitToStand} & \multicolumn{2}{c}{revolveDoor} \\
  stand & \cellcolor[gray]{0.9}{0.24} & \cellcolor[gray]{0.9}{0.06} & 0.00 & 0.00 & 0.00 & 0.00 & 0.00 & 0.00 & 0.00 & 0.00 & 0.83 & 0.00 & 0.00 & 0.00 & 0.00 & 0.00 \\ 
  walk & 0.02 & 0.01 & \cellcolor[gray]{0.9}{0.82} & \cellcolor[gray]{0.9}{0.91} & 0.00 & 0.17 & 0.39 & 0.50 & 0.00 & 0.30 & 0.00 & 0.00 & 0.15 & 0.30 & 0.00 & 0.10 \\ 
  stairUp & 0.04 & 0.04 & 0.12 & 0.05 & \cellcolor[gray]{0.9}{0.84} & \cellcolor[gray]{0.9}{0.78} & 0.16 & 0.01 & 0.00 & 0.00 & 0.13 & 0.13 & 0.50 & 0.70 & 0.43 & 0.90 \\ 
  stairDown & 0.00 & 0.02 & 0.05 & 0.04 & 0.11 & 0.04 & \cellcolor[gray]{0.9}{0.45} & \cellcolor[gray]{0.9}{0.48} & 0.00 & 0.25 & 0.00 & 0.00 & 0.10 & 0.00 & 0.13 & 0.00 \\ 
  standToSit & 0.00 & 0.00 & 0.01 & 0.00 & 0.00 & 0.00 & 0.00 & 0.00 & \cellcolor[gray]{0.9}{0.75} & \cellcolor[gray]{0.9}{0.00} & 0.00 & 0.00 & 0.25 & 0.00 & 0.43 & 0.00 \\ 
  sit & 0.69 & 0.75 & 0.00 & 0.00 & 0.00 & 0.00 & 0.00 & 0.00 & 0.25 & 0.30 & \cellcolor[gray]{0.9}{0.03} & \cellcolor[gray]{0.9}{0.87} & 0.00 & 0.00 & 0.00 & 0.00 \\ 
  sitToStand & 0.01 & 0.13 & 0.00 & 0.00 & 0.05 & 0.02 & 0.00 & 0.00 & 0.00 & 0.15 & 0.00 & 0.00 & \cellcolor[gray]{0.9}{0.00} & \cellcolor[gray]{0.9}{0.00} & 0.00 & 0.00 \\ 
  \tabularnewline
  \hline
  
    & \multicolumn{16}{c}{\bf{(c) Participant 3 }} \\%(0.71, 0.51; 0.80, 0.71)
    \hline
  \tabularnewline
  & \multicolumn{2}{c}{stand} & \multicolumn{2}{c}{walk} & \multicolumn{2}{c}{stairUp} & \multicolumn{2}{c}{stairDown} & \multicolumn{2}{c}{standToSit} & \multicolumn{2}{c}{sit} & \multicolumn{2}{c}{sitToStand} & \multicolumn{2}{c}{revolveDoor} \\
  stand & \cellcolor[gray]{0.9}{0.55} & \cellcolor[gray]{0.9}{0.73} & 0.01 & 0.00 & 0.00 & 0.00 & 0.00 & 0.00 & 0.00 & 0.00 & 0.12 & 0.21 & 0.00 & 0.00 & 0.10 & 0.00 \\ 
  walk & 0.00 & 0.00 & \cellcolor[gray]{0.9}{0.92} & \cellcolor[gray]{0.9}{0.94} & 0.01 & 0.08 & 0.26 & 0.41 & 0.00 & 0.45 & 0.00 & 0.00 & 0.25 & 0.60 & 0.00 & 0.00 \\ 
  stairUp & 0.12 & 0.06 & 0.06 & 0.04 & \cellcolor[gray]{0.9}{0.99} & \cellcolor[gray]{0.9}{0.86} & 0.25 & 0.26 & 0.10 & 0.00 & 0.02 & 0.07 & 0.35 & 0.40 & 0.90 & 1.00 \\ 
  stairDown & 0.00 & 0.02 & 0.01 & 0.01 & 0.00 & 0.05 & \cellcolor[gray]{0.9}{0.49} & \cellcolor[gray]{0.9}{0.32} & 0.00 & 0.50 & 0.00 & 0.00 & 0.00 & 0.00 & 0.00 & 0.00 \\ 
  standToSit & 0.03 & 0.13 & 0.00 & 0.01 & 0.00 & 0.00 & 0.00 & 0.01 & \cellcolor[gray]{0.9}{0.85} & \cellcolor[gray]{0.9}{0.00} & 0.01 & 0.01 & 0.00 & 0.00 & 0.00 & 0.00 \\ 
  sit & 0.30 & 0.06 & 0.00 & 0.00 & 0.00 & 0.00 & 0.00 & 0.00 & 0.00 & 0.00 & \cellcolor[gray]{0.9}{0.77} & \cellcolor[gray]{0.9}{0.69} & 0.00 & 0.00 & 0.00 & 0.00 \\ 
  sitToStand & 0.00 & 0.00 & 0.00 & 0.00 & 0.00 & 0.00 & 0.00 & 0.00 & 0.05 & 0.05 & 0.08 & 0.02 & \cellcolor[gray]{0.9}{0.40} & \cellcolor[gray]{0.9}{0.00} & 0.00 & 0.00 \\
  \tabularnewline
  \hline
  
    & \multicolumn{16}{c}{\bf{(d) Participant 4 }} \\%(0.58, 0.44; 0.68, 0.67)
    \hline
  \tabularnewline
  & \multicolumn{2}{c}{stand} & \multicolumn{2}{c}{walk} & \multicolumn{2}{c}{stairUp} & \multicolumn{2}{c}{stairDown} & \multicolumn{2}{c}{standToSit} & \multicolumn{2}{c}{sit} & \multicolumn{2}{c}{sitToStand} & \multicolumn{2}{c}{revolveDoor} \\
  stand & \cellcolor[gray]{0.9}{0.59} & \cellcolor[gray]{0.9}{0.52} & 0.00 & 0.00 & 0.00 & 0.00 & 0.06 & 0.00 & 0.32 & 0.11 & 0.17 & 0.24 & 0.25 & 0.00 & 0.23 & 0.00 \\ 
  walk & 0.00 & 0.04 & \cellcolor[gray]{0.9}{0.91} & \cellcolor[gray]{0.9}{0.81} & 0.34 & 0.05 & 0.39 & 0.41 & 0.00 & 0.00 & 0.00 & 0.00 & 0.00 & 0.30 & 0.57 & 0.00 \\ 
  stairUp & 0.00 & 0.04 & 0.06 & 0.13 & \cellcolor[gray]{0.9}{0.66} & \cellcolor[gray]{0.9}{0.84} & 0.06 & 0.21 & 0.00 & 0.89 & 0.00 & 0.03 & 0.70 & 0.50 & 0.00 & 0.33 \\ 
  stairDown & 0.04 & 0.03 & 0.01 & 0.05 & 0.00 & 0.11 & \cellcolor[gray]{0.9}{0.46} & \cellcolor[gray]{0.9}{0.36} & 0.00 & 0.00 & 0.00 & 0.00 & 0.05 & 0.00 & 0.00 & 0.00 \\ 
  standToSit & 0.09 & 0.06 & 0.00 & 0.00 & 0.00 & 0.00 & 0.01 & 0.02 & \cellcolor[gray]{0.9}{0.68} & \cellcolor[gray]{0.9}{0.00} & 0.00 & 0.10 & 0.00 & 0.20 & 0.20 & 0.67 \\ 
  sit & 0.29 & 0.25 & 0.00 & 0.00 & 0.00 & 0.00 & 0.00 & 0.00 & 0.00 & 0.00 & \cellcolor[gray]{0.9}{0.73} & \cellcolor[gray]{0.9}{0.56} & 0.00 & 0.00 & 0.00 & 0.00 \\ 
  sitToStand & 0.00 & 0.07 & 0.01 & 0.00 & 0.00 & 0.00 & 0.02 & 0.00 & 0.00 & 0.00 & 0.10 & 0.07 & \cellcolor[gray]{0.9}{0.00} & \cellcolor[gray]{0.9}{0.00} & 0.00 & 0.00 \\
  \hline
  \end{tabular}
  }
  \end{table}
  
  \begin{table}
\caption{True Activity Label versus Predicted Activity Label for Front Pocket Smartphone Accelerometer. \textit{See Table \ref{table:results_frontgyro} for details. The average of the diagonal elements for each data type (tri-axial vs. magnitude) is 0.63 vs. 0.73 for Participant 1, 0.52 vs. 0.57 for Participant 2, 0.50 vs. 0.66 for Participant 3, and 0.44 vs. 0.49 for Participant 4.}
\label{table:results_frontacc}}
\centering
\resizebox{\columnwidth}{!}{%
\small\addtolength{\tabcolsep}{-2pt}
\begin{tabular}{rllllllllllllllll}
  \hline
  & \multicolumn{16}{c}{\bf{(a) Participant 1 }} \\ %(0.63, 0.73; 0.60, 0.77)
    \hline
  \tabularnewline
  & \multicolumn{2}{c}{stand} & \multicolumn{2}{c}{walk} & \multicolumn{2}{c}{stairUp} & \multicolumn{2}{c}{stairDown} & \multicolumn{2}{c}{standToSit} & \multicolumn{2}{c}{sit} & \multicolumn{2}{c}{sitToStand} & \multicolumn{2}{c}{revolveDoor} \\
  stand & \cellcolor[gray]{0.9}{0.91} & \cellcolor[gray]{0.9}{0.85} & 0.00 & 0.00 & 0.00 & 0.00 & 0.01 & 0.00 & 0.00 & 0.00 & 0.00 & 0.00 & 0.00 & 0.33 & 0.14 & 0.00 \\ 
  walk & 0.01 & 0.00 & \cellcolor[gray]{0.9}{0.64} & \cellcolor[gray]{0.9}{0.62} & 0.00 & 0.07 & 0.05 & 0.12 & 0.00 & 0.00 & 0.00 & 0.00 & 0.00 & 0.00 & 0.00 & 0.00 \\ 
  stairUp & 0.07 & 0.03 & 0.18 & 0.26 & \cellcolor[gray]{0.9}{0.74} & \cellcolor[gray]{0.9}{0.92} & 0.52 & 0.10 & 0.00 & 0.00 & 0.00 & 0.00 & 0.38 & 0.24 & 0.46 & 0.68 \\ 
  stairDown & 0.00 & 0.00 & 0.18 & 0.10 & 0.26 & 0.00 & \cellcolor[gray]{0.9}{0.41} & \cellcolor[gray]{0.9}{0.78} & 0.00 & 0.00 & 0.00 & 0.00 & 0.14 & 0.00 & 0.40 & 0.00 \\ 
  standToSit & 0.00 & 0.03 & 0.00 & 0.02 & 0.00 & 0.01 & 0.00 & 0.00 & \cellcolor[gray]{0.9}{0.40} & \cellcolor[gray]{0.9}{0.85} & 0.00 & 0.00 & 0.00 & 0.24 & 0.00 & 0.32 \\ 
  sit & 0.00 & 0.09 & 0.00 & 0.00 & 0.00 & 0.00 & 0.00 & 0.00 & 0.60 & 0.15 & \cellcolor[gray]{0.9}{0.99} & \cellcolor[gray]{0.9}{0.98} & 0.14 & 0.10 & 0.00 & 0.00 \\ 
  sitToStand & 0.00 & 0.00 & 0.00 & 0.00 & 0.00 & 0.00 & 0.00 & 0.00 & 0.00 & 0.00 & 0.01 & 0.02 & \cellcolor[gray]{0.9}{0.33} & \cellcolor[gray]{0.9}{0.10} & 0.00 & 0.00 \\ 
  \tabularnewline
  \hline
  
    & \multicolumn{16}{c}{\bf{(b) Participant 2}} \\%(0.52, 0.57; 0.54, 0.61)
    \hline
  \tabularnewline
  & \multicolumn{2}{c}{stand} & \multicolumn{2}{c}{walk} & \multicolumn{2}{c}{stairUp} & \multicolumn{2}{c}{stairDown} & \multicolumn{2}{c}{standToSit} & \multicolumn{2}{c}{sit} & \multicolumn{2}{c}{sitToStand} & \multicolumn{2}{c}{revolveDoor} \\
  stand & \cellcolor[gray]{0.9}{1.00} & \cellcolor[gray]{0.9}{0.66} & 0.00 & 0.00 & 0.19 & 0.00 & 0.00 & 0.00 & 0.00 & 0.00 & 0.01 & 0.00 & 0.20 & 0.00 & 0.00 & 0.00 \\ 
  walk & 0.00 & 0.00 & \cellcolor[gray]{0.9}{0.80} & \cellcolor[gray]{0.9}{0.92} & 0.20 & 0.45 & 0.32 & 0.40 & 0.00 & 0.00 & 0.00 & 0.00 & 0.30 & 0.05 & 0.00 & 0.00 \\ 
  stairUp & 0.00 & 0.00 & 0.02 & 0.06 & \cellcolor[gray]{0.9}{0.28} & \cellcolor[gray]{0.9}{0.38} & 0.07 & 0.00 & 0.00 & 0.00 & 0.01 & 0.00 & 0.50 & 0.30 & 0.00 & 0.00 \\ 
  stairDown & 0.00 & 0.00 & 0.10 & 0.00 & 0.22 & 0.07 & \cellcolor[gray]{0.9}{0.53} & \cellcolor[gray]{0.9}{0.53} & 0.00 & 0.00 & 0.00 & 0.00 & 0.00 & 0.00 & 0.00 & 0.00 \\ 
  standToSit & 0.00 & 0.01 & 0.01 & 0.00 & 0.00 & 0.02 & 0.00 & 0.05 & \cellcolor[gray]{0.9}{1.00} & \cellcolor[gray]{0.9}{0.00} & 0.95 & 0.00 & 0.00 & 0.00 & 0.00 & 0.00 \\ 
  sit & 0.00 & 0.33 & 0.06 & 0.01 & 0.07 & 0.00 & 0.08 & 0.00 & 0.00 & 0.95 & \cellcolor[gray]{0.9}{0.00} & \cellcolor[gray]{0.9}{0.94} & 0.00 & 0.05 & 1.00 & 0.97 \\ 
  sitToStand & 0.00 & 0.00 & 0.00 & 0.01 & 0.04 & 0.08 & 0.00 & 0.02 & 0.00 & 0.05 & 0.03 & 0.06 & \cellcolor[gray]{0.9}{0.00} & \cellcolor[gray]{0.9}{0.60} & 0.00 & 0.03 \\
  \tabularnewline
  \hline
  
    & \multicolumn{16}{c}{\bf{(c) Participant 3 }} \\ %(0.50, 0.66; 0.60, 0.72)
    \hline
  \tabularnewline
  & \multicolumn{2}{c}{stand} & \multicolumn{2}{c}{walk} & \multicolumn{2}{c}{stairUp} & \multicolumn{2}{c}{stairDown} & \multicolumn{2}{c}{standToSit} & \multicolumn{2}{c}{sit} & \multicolumn{2}{c}{sitToStand} & \multicolumn{2}{c}{revolveDoor} \\
  stand & \cellcolor[gray]{0.9}{0.55} & \cellcolor[gray]{0.9}{0.70} & 0.00 & 0.00 & 0.00 & 0.00 & 0.00 & 0.00 & 0.00 & 0.00 & 0.00 & 0.00 & 0.20 & 0.00 & 0.00 & 0.00 \\ 
  walk & 0.00 & 0.01 & \cellcolor[gray]{0.9}{0.57} & \cellcolor[gray]{0.9}{0.62} & 0.03 & 0.03 & 0.07 & 0.27 & 0.05 & 0.00 & 0.00 & 0.00 & 0.30 & 0.30 & 0.00 & 0.00 \\ 
  stairUp & 0.41 & 0.02 & 0.30 & 0.20 & \cellcolor[gray]{0.9}{0.86} & \cellcolor[gray]{0.9}{0.97} & 0.50 & 0.17 & 0.40 & 0.10 & 0.02 & 0.02 & 0.50 & 0.70 & 0.80 & 0.05 \\ 
  stairDown & 0.01 & 0.04 & 0.13 & 0.15 & 0.06 & 0.00 & \cellcolor[gray]{0.9}{0.36} & \cellcolor[gray]{0.9}{0.56} & 0.00 & 0.00 & 0.00 & 0.00 & 0.00 & 0.00 & 0.20 & 0.00 \\ 
  standToSit & 0.00 & 0.05 & 0.00 & 0.02 & 0.04 & 0.00 & 0.00 & 0.00 & \cellcolor[gray]{0.9}{0.20} & \cellcolor[gray]{0.9}{0.90} & 0.00 & 0.11 & 0.00 & 0.00 & 0.00 & 0.95 \\ 
  sit & 0.03 & 0.17 & 0.00 & 0.00 & 0.01 & 0.00 & 0.06 & 0.00 & 0.35 & 0.00 & \cellcolor[gray]{0.9}{0.95} & \cellcolor[gray]{0.9}{0.87} & 0.00 & 0.00 & 0.00 & 0.00 \\ 
  sitToStand & 0.00 & 0.00 & 0.00 & 0.00 & 0.00 & 0.00 & 0.00 & 0.00 & 0.00 & 0.00 & 0.03 & 0.00 & \cellcolor[gray]{0.9}{0.00} & \cellcolor[gray]{0.9}{0.00} & 0.00 & 0.00 \\ 
  \tabularnewline
  \hline
  
    & \multicolumn{16}{c}{\bf{(d) Participant 4}} \\%(0.44, 0.49; 0.61, 0.61)
    \hline
  \tabularnewline
  & \multicolumn{2}{c}{stand} & \multicolumn{2}{c}{walk} & \multicolumn{2}{c}{stairUp} & \multicolumn{2}{c}{stairDown} & \multicolumn{2}{c}{standToSit} & \multicolumn{2}{c}{sit} & \multicolumn{2}{c}{sitToStand} & \multicolumn{2}{c}{revolveDoor} \\
  stand & \cellcolor[gray]{0.9}{0.04} & \cellcolor[gray]{0.9}{0.79} & 0.00 & 0.00 & 0.00 & 0.00 & 0.00 & 0.00 & 0.25 & 0.45 & 0.03 & 0.11 & 0.00 & 0.00 & 0.00 & 0.00 \\ 
  walk & 0.00 & 0.00 & \cellcolor[gray]{0.9}{0.86} & \cellcolor[gray]{0.9}{0.78} & 0.08 & 0.28 & 0.21 & 0.40 & 0.15 & 0.00 & 0.00 & 0.00 & 0.30 & 0.20 & 0.13 & 0.40 \\ 
  stairUp & 0.00 & 0.08 & 0.03 & 0.20 & \cellcolor[gray]{0.9}{0.41} & \cellcolor[gray]{0.9}{0.66} & 0.17 & 0.22 & 0.10 & 0.00 & 0.00 & 0.00 & 0.00 & 0.50 & 0.00 & 0.10 \\ 
  stairDown & 0.10 & 0.08 & 0.07 & 0.01 & 0.29 & 0.05 & \cellcolor[gray]{0.9}{0.56} & \cellcolor[gray]{0.9}{0.38} & 0.00 & 0.00 & 0.00 & 0.00 & 0.00 & 0.00 & 0.23 & 0.00 \\ 
  standToSit & 0.00 & 0.00 & 0.00 & 0.01 & 0.00 & 0.01 & 0.00 & 0.00 & \cellcolor[gray]{0.9}{0.30} & \cellcolor[gray]{0.9}{0.05} & 0.01 & 0.06 & 0.20 & 0.30 & 0.00 & 0.40 \\ 
  sit & 0.85 & 0.01 & 0.04 & 0.00 & 0.21 & 0.00 & 0.07 & 0.00 & 0.20 & 0.05 & \cellcolor[gray]{0.9}{0.88} & \cellcolor[gray]{0.9}{0.76} & 0.50 & 0.00 & 0.63 & 0.00 \\ 
  sitToStand & 0.00 & 0.03 & 0.00 & 0.00 & 0.00 & 0.00 & 0.00 & 0.00 & 0.00 & 0.45 & 0.08 & 0.07 & \cellcolor[gray]{0.9}{0.00} & \cellcolor[gray]{0.9}{0.00} & 0.00 & 0.10 \\ 
  \hline
  \end{tabular}
  }
  \end{table}
  
\begin{table}
\caption{True Activity Label versus Predicted Activity Label for Back Pocket Smartphone Accelerometer. \textit{See Table \ref{table:results_frontgyro} for details. The average of the diagonal elements for each data type (tri-axial vs. magnitude) is 0.70 vs. 0.72 for Participant 1, 0.07 vs. 0.56 for Participant 2, 0.65 vs. 0.55 for Participant 3, and 0.38 vs. 0.56 for Participant 4.}
\label{table:results_backacc}}
\centering
\resizebox{\columnwidth}{!}{%
\small\addtolength{\tabcolsep}{-2pt}
\begin{tabular}{rllllllllllllllll}
  \hline
  & \multicolumn{16}{c}{\bf{(a) Participant 1 }} \\%(0.70, 0.72; 0.69, 0.79)
    \hline
  \tabularnewline
  & \multicolumn{2}{c}{stand} & \multicolumn{2}{c}{walk} & \multicolumn{2}{c}{stairUp} & \multicolumn{2}{c}{stairDown} & \multicolumn{2}{c}{standToSit} & \multicolumn{2}{c}{sit} & \multicolumn{2}{c}{sitToStand} & \multicolumn{2}{c}{revolveDoor} \\
  stand & \cellcolor[gray]{0.9}{0.93} & \cellcolor[gray]{0.9}{0.71} & 0.01 & 0.00 & 0.03 & 0.00 & 0.00 & 0.00 & 0.00 & 0.10 & 0.00 & 0.13 & 0.20 & 0.00 & 0.38 & 0.00 \\ 
  walk & 0.00 & 0.00 & \cellcolor[gray]{0.9}{0.71} & \cellcolor[gray]{0.9}{0.76} & 0.03 & 0.02 & 0.21 & 0.07 & 0.00 & 0.00 & 0.00 & 0.00 & 0.00 & 0.20 & 0.04 & 0.08 \\ 
  stairUp & 0.00 & 0.01 & 0.21 & 0.18 & \cellcolor[gray]{0.9}{0.74} & \cellcolor[gray]{0.9}{0.95} & 0.14 & 0.28 & 0.00 & 0.00 & 0.00 & 0.00 & 0.00 & 0.00 & 0.34 & 0.50 \\ 
  stairDown & 0.00 & 0.00 & 0.06 & 0.03 & 0.10 & 0.00 & \cellcolor[gray]{0.9}{0.62} & \cellcolor[gray]{0.9}{0.65} & 0.00 & 0.00 & 0.00 & 0.00 & 0.00 & 0.00 & 0.00 & 0.00 \\ 
  standToSit & 0.00 & 0.04 & 0.01 & 0.03 & 0.00 & 0.03 & 0.01 & 0.00 & \cellcolor[gray]{0.9}{0.55} & \cellcolor[gray]{0.9}{0.90} & 0.05 & 0.03 & 0.00 & 0.40 & 0.06 & 0.32 \\ 
  sit & 0.06 & 0.23 & 0.01 & 0.00 & 0.10 & 0.00 & 0.01 & 0.00 & 0.00 & 0.00 & \cellcolor[gray]{0.9}{0.91} & \cellcolor[gray]{0.9}{0.84} & 0.35 & 0.15 & 0.18 & 0.10 \\ 
  sitToStand & 0.01 & 0.00 & 0.00 & 0.00 & 0.00 & 0.00 & 0.00 & 0.00 & 0.45 & 0.00 & 0.04 & 0.00 & \cellcolor[gray]{0.9}{0.45} & \cellcolor[gray]{0.9}{0.25} & 0.00 & 0.00 \\ 
  \tabularnewline
  \hline
  
    & \multicolumn{16}{c}{\bf{(b) Participant 2 }} \\%(0.07, 0.56; 0.14, 0.60)
    \hline
  \tabularnewline
  & \multicolumn{2}{c}{stand} & \multicolumn{2}{c}{walk} & \multicolumn{2}{c}{stairUp} & \multicolumn{2}{c}{stairDown} & \multicolumn{2}{c}{standToSit} & \multicolumn{2}{c}{sit} & \multicolumn{2}{c}{sitToStand} & \multicolumn{2}{c}{revolveDoor} \\
  stand & \cellcolor[gray]{0.9}{0.03} & \cellcolor[gray]{0.9}{0.80} & 0.00 & 0.00 & 0.00 & 0.00 & 0.00 & 0.00 & 0.05 & 0.05 & 0.51 & 0.09 & 0.00 & 0.00 & 0.00 & 0.00 \\ 
  walk & 0.00 & 0.04 & \cellcolor[gray]{0.9}{0.36} & \cellcolor[gray]{0.9}{0.93} & 0.03 & 0.14 & 0.16 & 0.49 & 0.00 & 0.20 & 0.00 & 0.00 & 0.00 & 0.15 & 0.00 & 0.00 \\ 
  stairUp & 0.00 & 0.00 & 0.02 & 0.02 & \cellcolor[gray]{0.9}{0.00} & \cellcolor[gray]{0.9}{0.41} & 0.11 & 0.06 & 0.00 & 0.15 & 0.00 & 0.02 & 0.00 & 0.60 & 0.00 & 0.00 \\ 
  stairDown & 0.00 & 0.00 & 0.04 & 0.01 & 0.00 & 0.28 & \cellcolor[gray]{0.9}{0.06} & \cellcolor[gray]{0.9}{0.46} & 0.00 & 0.00 & 0.00 & 0.00 & 0.00 & 0.00 & 0.00 & 0.00 \\ 
  standToSit & 0.02 & 0.00 & 0.16 & 0.02 & 0.04 & 0.08 & 0.05 & 0.00 & \cellcolor[gray]{0.9}{0.00} & \cellcolor[gray]{0.9}{0.45} & 0.07 & 0.06 & 0.95 & 0.25 & 0.00 & 0.67 \\ 
  sit & 0.25 & 0.16 & 0.35 & 0.00 & 0.68 & 0.06 & 0.55 & 0.00 & 0.20 & 0.15 & \cellcolor[gray]{0.9}{0.04} & \cellcolor[gray]{0.9}{0.83} & 0.05 & 0.00 & 1.00 & 0.33 \\ 
  sitToStand & 0.71 & 0.00 & 0.06 & 0.01 & 0.24 & 0.03 & 0.07 & 0.00 & 0.75 & 0.00 & 0.38 & 0.00 & \cellcolor[gray]{0.9}{0.00} & \cellcolor[gray]{0.9}{0.00} & 0.00 & 0.00 \\ 
  \tabularnewline
  \hline
  
    & \multicolumn{16}{c}{\bf{(c) Participant 3 }} \\%(0.65, 0.55; 0.58, 0.68)
    \hline
  \tabularnewline
  & \multicolumn{2}{c}{stand} & \multicolumn{2}{c}{walk} & \multicolumn{2}{c}{stairUp} & \multicolumn{2}{c}{stairDown} & \multicolumn{2}{c}{standToSit} & \multicolumn{2}{c}{sit} & \multicolumn{2}{c}{sitToStand} & \multicolumn{2}{c}{revolveDoor} \\
  stand & \cellcolor[gray]{0.9}{0.84} & \cellcolor[gray]{0.9}{0.87} & 0.00 & 0.00 & 0.01 & 0.00 & 0.04 & 0.00 & 0.00 & 0.15 & 0.00 & 0.10 & 0.00 & 0.00 & 0.00 & 0.20 \\ 
  walk & 0.00 & 0.03 & \cellcolor[gray]{0.9}{0.67} & \cellcolor[gray]{0.9}{0.65} & 0.04 & 0.04 & 0.18 & 0.21 & 0.00 & 0.00 & 0.00 & 0.00 & 0.00 & 0.05 & 0.00 & 0.00 \\ 
  stairUp & 0.01 & 0.03 & 0.24 & 0.13 & \cellcolor[gray]{0.9}{0.71} & \cellcolor[gray]{0.9}{0.92} & 0.42 & 0.26 & 0.00 & 0.75 & 0.00 & 0.01 & 0.55 & 0.75 & 1.00 & 0.55 \\ 
  stairDown & 0.10 & 0.04 & 0.08 & 0.19 & 0.22 & 0.02 & \cellcolor[gray]{0.9}{0.36} & \cellcolor[gray]{0.9}{0.49} & 0.00 & 0.00 & 0.00 & 0.00 & 0.15 & 0.00 & 0.00 & 0.00 \\ 
  standToSit & 0.05 & 0.03 & 0.01 & 0.02 & 0.01 & 0.02 & 0.00 & 0.04 & \cellcolor[gray]{0.9}{0.70} & \cellcolor[gray]{0.9}{0.10} & 0.00 & 0.04 & 0.00 & 0.20 & 0.00 & 0.00 \\ 
  sit & 0.00 & 0.00 & 0.00 & 0.01 & 0.00 & 0.00 & 0.00 & 0.00 & 0.30 & 0.00 & \cellcolor[gray]{0.9}{0.98} & \cellcolor[gray]{0.9}{0.83} & 0.00 & 0.00 & 0.00 & 0.25 \\ 
  sitToStand & 0.00 & 0.00 & 0.00 & 0.00 & 0.00 & 0.00 & 0.00 & 0.00 & 0.00 & 0.00 & 0.02 & 0.01 & \cellcolor[gray]{0.9}{0.30} & \cellcolor[gray]{0.9}{0.00} & 0.00 & 0.00 \\ 
  \tabularnewline
  \hline
  
    & \multicolumn{16}{c}{\bf{(d) Participant 4 }} \\%(0.38, 0.56; 0.65, 0.63)
    \hline
  \tabularnewline
  & \multicolumn{2}{c}{stand} & \multicolumn{2}{c}{walk} & \multicolumn{2}{c}{stairUp} & \multicolumn{2}{c}{stairDown} & \multicolumn{2}{c}{standToSit} & \multicolumn{2}{c}{sit} & \multicolumn{2}{c}{sitToStand} & \multicolumn{2}{c}{revolveDoor} \\
  stand & \cellcolor[gray]{0.9}{0.24} & \cellcolor[gray]{0.9}{0.82} & 0.00 & 0.01 & 0.00 & 0.01 & 0.00 & 0.00 & 0.20 & 0.05 & 0.87 & 0.24 & 0.00 & 0.00 & 0.00 & 0.00 \\ 
  walk & 0.00 & 0.00 & \cellcolor[gray]{0.9}{0.58} & \cellcolor[gray]{0.9}{0.70} & 0.10 & 0.21 & 0.23 & 0.38 & 0.40 & 0.40 & 0.00 & 0.00 & 0.00 & 0.40 & 0.00 & 0.23 \\ 
  stairUp & 0.03 & 0.01 & 0.31 & 0.27 & \cellcolor[gray]{0.9}{0.69} & \cellcolor[gray]{0.9}{0.62} & 0.08 & 0.04 & 0.00 & 0.00 & 0.00 & 0.00 & 0.05 & 0.45 & 0.63 & 0.50 \\ 
  stairDown & 0.07 & 0.10 & 0.08 & 0.00 & 0.16 & 0.11 & \cellcolor[gray]{0.9}{0.69} & \cellcolor[gray]{0.9}{0.58} & 0.00 & 0.00 & 0.00 & 0.00 & 0.20 & 0.00 & 0.00 & 0.00 \\ 
  standToSit & 0.12 & 0.03 & 0.02 & 0.02 & 0.04 & 0.05 & 0.00 & 0.00 & \cellcolor[gray]{0.9}{0.40} & \cellcolor[gray]{0.9}{0.55} & 0.04 & 0.02 & 0.70 & 0.15 & 0.37 & 0.27 \\ 
  sit & 0.52 & 0.04 & 0.00 & 0.00 & 0.00 & 0.00 & 0.00 & 0.00 & 0.00 & 0.00 & \cellcolor[gray]{0.9}{0.00} & \cellcolor[gray]{0.9}{0.68} & 0.00 & 0.00 & 0.00 & 0.00 \\ 
  sitToStand & 0.02 & 0.00 & 0.01 & 0.00 & 0.02 & 0.00 & 0.00 & 0.00 & 0.00 & 0.00 & 0.08 & 0.06 & \cellcolor[gray]{0.9}{0.05} & \cellcolor[gray]{0.9}{0.00} & 0.00 & 0.00 \\ 
  \hline
  \end{tabular}
  }
  \end{table}
  
% JPO: here to need to explain the numbers after Participant 1 (0.49, etc.) 
\begin{table}
\caption{Distribution of Activity Label Predicted Using Tri-axial Data, for Each Walking Speed. \textit{These results are for the test segment where the participant walked at different speeds. In each subtable, the first column heading indicates the walking speed (Slow, Normal, Fast). The second heading indicates the sensor type, either accelerometer (``Acc'') or gyroscope (``Gyro''). The third heading indicates whether the sensor was in the front or back pocket phone. The rows are the predicted activity labels. Each column of values shows the distribution of the predicted activity labels and sums to 1. The shaded row highlights the proportion of labels correctly predicted as walking. The average of the shaded elements for each data type (Acc-front, Acc-back, Gyro-front, Gyro-back) is (0.49, 0.77, 0.69, 0.91) for Participant 1,  (0.72, 0.33, 0.89, 0.53) for Participant 2,  (0.48, 0.33, 0.56, 0.71) for Participant 3, and (0.54, 0.27, 0.80, 0.73) for Participant 4.} \label{table:walking_speeds_TRI}}
\centering
\scalebox{0.8}{
\small\addtolength{\tabcolsep}{-2pt}
\begin{tabular}{rllllllllllll}
  \hline
  & \multicolumn{12}{c}{\bf{(a) Participant 1 }} \\%(0.49, 0.77, 0.69, 0.91)
    \hline
 & \multicolumn{4}{c}{Slow} & \multicolumn{4}{c}{Normal} & \multicolumn{4}{c}{Fast}\\
\cline{2-5} \cline{6-10} \cline{11-13}
  & \multicolumn{2}{c}{Acc} & \multicolumn{2}{c}{Gyro} & \multicolumn{2}{c}{Acc} & \multicolumn{2}{c}{Gyro} & \multicolumn{2}{c}{Acc} & \multicolumn{2}{c}{Gyro}\\
 & front & back & front & back & front & back & front & back & front & back & front & back \\ 
  \hline
stand & 0.01 & 0.00 & 0.00 & 0.00 & 0.02 & 0.00 & 0.00 & 0.00 & 0.00 & 0.00 & 0.00 & 0.00 \\ 
  walk & \cellcolor[gray]{0.9}{0.59} & \cellcolor[gray]{0.9}{0.60} & \cellcolor[gray]{0.9}{0.55} & \cellcolor[gray]{0.9}{0.80} & \cellcolor[gray]{0.9}{0.59} & \cellcolor[gray]{0.9}{0.88} & \cellcolor[gray]{0.9}{0.82} & \cellcolor[gray]{0.9}{0.93} & \cellcolor[gray]{0.9}{0.30} & \cellcolor[gray]{0.9}{0.84} & \cellcolor[gray]{0.9}{0.71} & \cellcolor[gray]{0.9}{1.00} \\ 
  stairUp & 0.29 & 0.29 & 0.45 & 0.18 & 0.17 & 0.06 & 0.17 & 0.07 & 0.21 & 0.08 & 0.11 & 0.00 \\ 
  stairDown & 0.11 & 0.08 & 0.00 & 0.02 & 0.21 & 0.05 & 0.01 & 0.00 & 0.49 & 0.08 & 0.18 & 0.00 \\ 
  standToSit & 0.00 & 0.00 & 0.00 & 0.00 & 0.00 & 0.00 & 0.00 & 0.00 & 0.00 & 0.00 & 0.00 & 0.00 \\ 
  sit & 0.00 & 0.03 & 0.00 & 0.00 & 0.00 & 0.00 & 0.00 & 0.00 & 0.00 & 0.00 & 0.00 & 0.00 \\ 
  sitToStand & 0.00 & 0.00 & 0.00 & 0.00 & 0.00 & 0.00 & 0.00 & 0.00 & 0.00 & 0.00 & 0.00 & 0.00 \\ 
  \tabularnewline
    \hline
  & \multicolumn{12}{c}{\bf{(b) Participant 2 }} \\%(0.72, 0.33, 0.89, 0.53)
    \hline
 & \multicolumn{4}{c}{Slow} & \multicolumn{4}{c}{Normal} & \multicolumn{4}{c}{Fast}\\
\cline{2-5} \cline{6-10} \cline{11-13}
  & \multicolumn{2}{c}{Acc} & \multicolumn{2}{c}{Gyro} & \multicolumn{2}{c}{Acc} & \multicolumn{2}{c}{Gyro} & \multicolumn{2}{c}{Acc} & \multicolumn{2}{c}{Gyro}\\
 & front & back & front & back & front & back & front & back & front & back & front & back \\ 
  \hline
  stand & 0.39 & 0.00 & 0.00 & 0.00 & 0.00 & 0.00 & 0.00 & 0.00 & 0.00 & 0.00 & 0.00 & 0.00 \\ 
  walk & \cellcolor[gray]{0.9}{0.24} & \cellcolor[gray]{0.9}{0.00} & \cellcolor[gray]{0.9}{0.67} & \cellcolor[gray]{0.9}{0.02} & \cellcolor[gray]{0.9}{1.00} & \cellcolor[gray]{0.9}{0.67} & \cellcolor[gray]{0.9}{1.00} & \cellcolor[gray]{0.9}{1.00} & \cellcolor[gray]{0.9}{0.91} & \cellcolor[gray]{0.9}{0.31} & \cellcolor[gray]{0.9}{1.00} & \cellcolor[gray]{0.9}{0.58} \\ 
  stairUp & 0.18 & 0.00 & 0.00 & 0.40 & 0.00 & 0.00 & 0.00 & 0.00 & 0.01 & 0.23 & 0.00 & 0.04 \\ 
  stairDown & 0.18 & 0.00 & 0.00 & 0.50 & 0.00 & 0.08 & 0.00 & 0.00 & 0.08 & 0.00 & 0.00 & 0.38 \\ 
  standToSit & 0.00 & 0.99 & 0.00 & 0.05 & 0.00 & 0.23 & 0.00 & 0.00 & 0.00 & 0.45 & 0.00 & 0.00 \\ 
  sit & 0.00 & 0.01 & 0.04 & 0.00 & 0.00 & 0.00 & 0.00 & 0.00 & 0.00 & 0.00 & 0.00 & 0.00 \\ 
  sitToStand & 0.00 & 0.00 & 0.29 & 0.03 & 0.00 & 0.02 & 0.00 & 0.00 & 0.00 & 0.01 & 0.00 & 0.00 \\
    \tabularnewline
    \hline
  & \multicolumn{12}{c}{\bf{(c) Participant 3 }} \\%(0.48, 0.33, 0.56, 0.71)
    \hline
 & \multicolumn{4}{c}{Slow} & \multicolumn{4}{c}{Normal} & \multicolumn{4}{c}{Fast}\\
\cline{2-5} \cline{6-10} \cline{11-13}
  & \multicolumn{2}{c}{Acc} & \multicolumn{2}{c}{Gyro} & \multicolumn{2}{c}{Acc} & \multicolumn{2}{c}{Gyro} & \multicolumn{2}{c}{Acc} & \multicolumn{2}{c}{Gyro}\\
 & front & back & front & back & front & back & front & back & front & back & front & back \\ 
  \hline
  stand & 0.00 & 0.00 & 0.00 & 0.00 & 0.00 & 0.00 & 0.00 & 0.00 & 0.00 & 0.00 & 0.00 & 0.00 \\ 
  walk & \cellcolor[gray]{0.9}{0.11} & \cellcolor[gray]{0.9}{0.00} & \cellcolor[gray]{0.9}{0.23} & \cellcolor[gray]{0.9}{0.22} & \cellcolor[gray]{0.9}{0.86} & \cellcolor[gray]{0.9}{0.74} & \cellcolor[gray]{0.9}{0.69} & \cellcolor[gray]{0.9}{0.98} & \cellcolor[gray]{0.9}{0.47} & \cellcolor[gray]{0.9}{0.23} & \cellcolor[gray]{0.9}{0.75} & \cellcolor[gray]{0.9}{0.93} \\ 
  stairUp & 0.85 & 0.86 & 0.72 & 0.78 & 0.12 & 0.21 & 0.22 & 0.02 & 0.51 & 0.11 & 0.23 & 0.01 \\ 
  stairDown & 0.04 & 0.01 & 0.05 & 0.00 & 0.02 & 0.03 & 0.08 & 0.00 & 0.01 & 0.66 & 0.03 & 0.06 \\ 
  standToSit & 0.00 & 0.02 & 0.00 & 0.00 & 0.00 & 0.02 & 0.00 & 0.00 & 0.00 & 0.00 & 0.00 & 0.00 \\ 
  sit & 0.00 & 0.00 & 0.00 & 0.00 & 0.00 & 0.00 & 0.00 & 0.00 & 0.00 & 0.00 & 0.00 & 0.00 \\ 
  sitToStand & 0.00 & 0.10 & 0.00 & 0.00 & 0.00 & 0.00 & 0.00 & 0.00 & 0.00 & 0.00 & 0.00 & 0.00 \\ 
    \tabularnewline
    \hline
 &  \multicolumn{12}{c}{\bf{(d) Participant 4 }} \\%(0.54, 0.27, 0.80, 0.73)
    \hline
 & \multicolumn{4}{c}{Slow} & \multicolumn{4}{c}{Normal} & \multicolumn{4}{c}{Fast}\\
\cline{2-5} \cline{6-10} \cline{11-13}
  & \multicolumn{2}{c}{Acc} & \multicolumn{2}{c}{Gyro} & \multicolumn{2}{c}{Acc} & \multicolumn{2}{c}{Gyro} & \multicolumn{2}{c}{Acc} & \multicolumn{2}{c}{Gyro}\\
 & front & back & front & back & front & back & front & back & front & back & front & back \\ 
  \hline
  stand & 0.00 & 0.00 & 0.01 & 0.04 & 0.00 & 0.00 & 0.00 & 0.00 & 0.00 & 0.00 & 0.00 & 0.00 \\ 
  walk & \cellcolor[gray]{0.9}{0.06} & \cellcolor[gray]{0.9}{0.04} & \cellcolor[gray]{0.9}{0.41} & \cellcolor[gray]{0.9}{0.49} & \cellcolor[gray]{0.9}{0.91} & \cellcolor[gray]{0.9}{0.55} & \cellcolor[gray]{0.9}{1.00} & \cellcolor[gray]{0.9}{1.00} & \cellcolor[gray]{0.9}{0.66} & \cellcolor[gray]{0.9}{0.24} & \cellcolor[gray]{0.9}{1.00} & \cellcolor[gray]{0.9}{0.69} \\ 
  stairUp & 0.05 & 0.69 & 0.02 & 0.46 & 0.01 & 0.42 & 0.00 & 0.00 & 0.02 & 0.40 & 0.00 & 0.31 \\ 
  stairDown & 0.21 & 0.25 & 0.56 & 0.00 & 0.08 & 0.03 & 0.00 & 0.00 & 0.32 & 0.36 & 0.00 & 0.00 \\ 
  standToSit & 0.00 & 0.01 & 0.00 & 0.00 & 0.00 & 0.00 & 0.00 & 0.00 & 0.00 & 0.00 & 0.00 & 0.00 \\ 
  sit & 0.69 & 0.00 & 0.00 & 0.00 & 0.01 & 0.00 & 0.00 & 0.00 & 0.00 & 0.00 & 0.00 & 0.00 \\ 
  sitToStand & 0.00 & 0.02 & 0.00 & 0.01 & 0.00 & 0.00 & 0.00 & 0.00 & 0.00 & 0.00 & 0.00 & 0.00 \\ 
   \hline
\end{tabular}}
\end{table}

\clearpage

\begin{table}
\caption{Distribution of Activity Label Predicted Using Magnitude Data, for Each Walking Speed \textit{See Table \ref{table:walking_speeds_TRI} for details. The average of the shaded elements for each data-type (acc-front, acc-back, gyro-front, gyro-back) is (0.38, 0.79, 0.72, 0.81) for Participant 1,  (0.68, 0.61, 0.64, 0.83) for Participant 2,  (0.34, 0.55, 0.48, 0.64) for Participant 3, and (0.48, 0.57, 0.53, 0.62) for Participant 4.}
 \label{table:walking_speeds_MAG}}
\centering
\scalebox{0.9}{
\small\addtolength{\tabcolsep}{-2pt}
\begin{tabular}{rllllllllllll}
  \hline
  & \multicolumn{12}{c}{\bf{(a) Participant 1 }} \\%(0.38, 0.79, 0.72, 0.81)
    \hline
 & \multicolumn{4}{c}{Slow} & \multicolumn{4}{c}{Normal} & \multicolumn{4}{c}{Fast}\\
\cline{2-5} \cline{6-10} \cline{11-13}
  & \multicolumn{2}{c}{Acc} & \multicolumn{2}{c}{Gyro} & \multicolumn{2}{c}{Acc} & \multicolumn{2}{c}{Gyro} & \multicolumn{2}{c}{Acc} & \multicolumn{2}{c}{Gyro}\\
 & front & back & front & back & front & back & front & back & front & back & front & back \\ 
  \hline
stand & 0.00 & 0.00 & 0.00 & 0.00 & 0.00 & 0.00 & 0.00 & 0.00 & 0.00 & 0.00 & 0.00 & 0.00 \\ 
  walk & \cellcolor[gray]{0.9}{0.27} & \cellcolor[gray]{0.9}{0.82} & \cellcolor[gray]{0.9}{0.22} & \cellcolor[gray]{0.9}{0.57} & \cellcolor[gray]{0.9}{0.52} & \cellcolor[gray]{0.9}{0.73} & \cellcolor[gray]{0.9}{0.94} & \cellcolor[gray]{0.9}{0.94} & \cellcolor[gray]{0.9}{0.34} & \cellcolor[gray]{0.9}{0.82} & \cellcolor[gray]{0.9}{1.00} & \cellcolor[gray]{0.9}{0.94} \\ 
  stairUp & 0.67 & 0.11 & 0.76 & 0.22 & 0.26 & 0.23 & 0.03 & 0.00 & 0.12 & 0.05 & 0.00 & 0.00 \\ 
  stairDown & 0.03 & 0.03 & 0.02 & 0.21 & 0.21 & 0.03 & 0.00 & 0.06 & 0.54 & 0.12 & 0.00 & 0.06 \\ 
  standToSit & 0.03 & 0.02 & 0.00 & 0.00 & 0.00 & 0.00 & 0.00 & 0.00 & 0.00 & 0.00 & 0.00 & 0.00 \\ 
  sit & 0.00 & 0.00 & 0.00 & 0.00 & 0.00 & 0.00 & 0.00 & 0.00 & 0.00 & 0.00 & 0.00 & 0.00 \\ 
  sitToStand & 0.00 & 0.03 & 0.00 & 0.00 & 0.00 & 0.00 & 0.02 & 0.00 & 0.00 & 0.00 & 0.00 & 0.00 \\ 
  \tabularnewline
    \hline
  & \multicolumn{12}{c}{\bf{(b) Participant 2 }} \\%(0.68, 0.61, 0.64, 0.83)
    \hline
 & \multicolumn{4}{c}{Slow} & \multicolumn{4}{c}{Normal} & \multicolumn{4}{c}{Fast}\\
\cline{2-5} \cline{6-10} \cline{11-13}
  & \multicolumn{2}{c}{Acc} & \multicolumn{2}{c}{Gyro} & \multicolumn{2}{c}{Acc} & \multicolumn{2}{c}{Gyro} & \multicolumn{2}{c}{Acc} & \multicolumn{2}{c}{Gyro}\\
 & front & back & front & back & front & back & front & back & front & back & front & back \\ 
  \hline
 stand & 0.00 & 0.00 & 0.00 & 0.00 & 0.00 & 0.00 & 0.00 & 0.00 & 0.00 & 0.00 & 0.00 & 0.00 \\ 
  walk & \cellcolor[gray]{0.9}{0.15} & \cellcolor[gray]{0.9}{0.08} & \cellcolor[gray]{0.9}{0.00} & \cellcolor[gray]{0.9}{0.49} & \cellcolor[gray]{0.9}{1.00} & \cellcolor[gray]{0.9}{1.00} & \cellcolor[gray]{0.9}{0.92} & \cellcolor[gray]{0.9}{1.00} & \cellcolor[gray]{0.9}{0.89} & \cellcolor[gray]{0.9}{0.74} & \cellcolor[gray]{0.9}{1.00} & \cellcolor[gray]{0.9}{1.00} \\ 
  stairUp & 0.83 & 0.26 & 0.98 & 0.51 & 0.00 & 0.00 & 0.08 & 0.00 & 0.01 & 0.16 & 0.00 & 0.00 \\ 
  stairDown & 0.00 & 0.00 & 0.00 & 0.00 & 0.00 & 0.00 & 0.00 & 0.00 & 0.05 & 0.10 & 0.00 & 0.00 \\ 
  standToSit & 0.00 & 0.39 & 0.00 & 0.00 & 0.00 & 0.00 & 0.00 & 0.00 & 0.05 & 0.00 & 0.00 & 0.00 \\ 
  sit & 0.02 & 0.00 & 0.00 & 0.00 & 0.00 & 0.00 & 0.00 & 0.00 & 0.00 & 0.00 & 0.00 & 0.00 \\ 
  sitToStand & 0.00 & 0.27 & 0.02 & 0.00 & 0.00 & 0.00 & 0.00 & 0.00 & 0.00 & 0.00 & 0.00 & 0.00 \\ 
    \tabularnewline
    \hline
  & \multicolumn{12}{c}{\bf{(c) Participant 3 }} \\%(0.34, 0.55, 0.48, 0.64)
    \hline
 & \multicolumn{4}{c}{Slow} & \multicolumn{4}{c}{Normal} & \multicolumn{4}{c}{Fast}\\
\cline{2-5} \cline{6-10} \cline{11-13}
  & \multicolumn{2}{c}{Acc} & \multicolumn{2}{c}{Gyro} & \multicolumn{2}{c}{Acc} & \multicolumn{2}{c}{Gyro} & \multicolumn{2}{c}{Acc} & \multicolumn{2}{c}{Gyro}\\
 & front & back & front & back & front & back & front & back & front & back & front & back \\ 
  \hline
 stand & 0.00 & 0.00 & 0.00 & 0.00 & 0.00 & 0.00 & 0.00 & 0.00 & 0.00 & 0.00 & 0.00 & 0.00 \\ 
  walk & \cellcolor[gray]{0.9}{0.06} & \cellcolor[gray]{0.9}{0.06} & \cellcolor[gray]{0.9}{0.00} & \cellcolor[gray]{0.9}{0.00} & \cellcolor[gray]{0.9}{0.56} & \cellcolor[gray]{0.9}{0.83} & \cellcolor[gray]{0.9}{0.47} & \cellcolor[gray]{0.9}{0.93} & \cellcolor[gray]{0.9}{0.39} & \cellcolor[gray]{0.9}{0.75} & \cellcolor[gray]{0.9}{0.96} & \cellcolor[gray]{0.9}{1.00} \\ 
  stairUp & 0.73 & 0.67 & 0.70 & 0.91 & 0.34 & 0.03 & 0.05 & 0.07 & 0.09 & 0.11 & 0.00 & 0.00 \\ 
  stairDown & 0.00 & 0.08 & 0.23 & 0.09 & 0.10 & 0.10 & 0.48 & 0.00 & 0.52 & 0.11 & 0.04 & 0.00 \\ 
  standToSit & 0.19 & 0.19 & 0.07 & 0.00 & 0.00 & 0.00 & 0.00 & 0.00 & 0.00 & 0.03 & 0.00 & 0.00 \\ 
  sit & 0.00 & 0.01 & 0.00 & 0.00 & 0.00 & 0.00 & 0.00 & 0.00 & 0.00 & 0.00 & 0.00 & 0.00 \\ 
  sitToStand & 0.02 & 0.00 & 0.00 & 0.00 & 0.00 & 0.03 & 0.00 & 0.00 & 0.00 & 0.00 & 0.00 & 0.00 \\ 
    \tabularnewline
    \hline
  & \multicolumn{12}{c}{\bf{(d) Participant 4 }} \\%(0.48, 0.57, 0.53, 0.62)
    \hline
 & \multicolumn{4}{c}{Slow} & \multicolumn{4}{c}{Normal} & \multicolumn{4}{c}{Fast}\\
\cline{2-5} \cline{6-10} \cline{11-13}
  & \multicolumn{2}{c}{Acc} & \multicolumn{2}{c}{Gyro} & \multicolumn{2}{c}{Acc} & \multicolumn{2}{c}{Gyro} & \multicolumn{2}{c}{Acc} & \multicolumn{2}{c}{Gyro}\\
 & front & back & front & back & front & back & front & back & front & back & front & back \\ 
  \hline
stand & 0.00 & 0.36 & 0.00 & 0.00 & 0.00 & 0.00 & 0.00 & 0.00 & 0.00 & 0.00 & 0.00 & 0.00 \\ 
  walk & \cellcolor[gray]{0.9}{0.34} & \cellcolor[gray]{0.9}{0.03} & \cellcolor[gray]{0.9}{0.28} & \cellcolor[gray]{0.9}{0.01} & \cellcolor[gray]{0.9}{0.65} & \cellcolor[gray]{0.9}{0.92} & \cellcolor[gray]{0.9}{0.99} & \cellcolor[gray]{0.9}{0.91} & \cellcolor[gray]{0.9}{0.45} & \cellcolor[gray]{0.9}{0.76} & \cellcolor[gray]{0.9}{0.34} & \cellcolor[gray]{0.9}{0.95} \\ 
  stairUp & 0.18 & 0.30 & 0.17 & 0.68 & 0.35 & 0.08 & 0.01 & 0.06 & 0.39 & 0.14 & 0.66 & 0.00 \\ 
  stairDown & 0.00 & 0.00 & 0.48 & 0.04 & 0.00 & 0.00 & 0.00 & 0.03 & 0.16 & 0.10 & 0.00 & 0.05 \\ 
  standToSit & 0.33 & 0.30 & 0.07 & 0.17 & 0.00 & 0.00 & 0.00 & 0.00 & 0.00 & 0.00 & 0.00 & 0.00 \\ 
  sit & 0.00 & 0.00 & 0.00 & 0.00 & 0.00 & 0.00 & 0.00 & 0.00 & 0.00 & 0.00 & 0.00 & 0.00 \\ 
  sitToStand & 0.14 & 0.01 & 0.01 & 0.10 & 0.00 & 0.00 & 0.00 & 0.00 & 0.00 & 0.00 & 0.00 & 0.00 \\ 
   \hline
\end{tabular}}
\end{table}

\clearpage

%\begin{table}[!p]
%\tblcaption{Transforming Gyroscope or Accelerometer Data to a Standard Frame of Reference (having the phone upside down and with the phone's face against the leg). For a given axis, the data is either left unchanged (indicated by 1) or multiplied by -1 (indicated by -1).
%\label{table:transform}}
%{\tabcolsep=4.25pt
%\begin{tabular}{@{}ccccc@{}}
%\tblhead{Face Against Leg? & Upside Down? & $x$ & $y$ & $z$ } 
%no & no & { }1 & -1 & -1 \\
%no & yes & -1 & { }1 & -1 \\
%yes & no & -1 & -1 & { }1  \\
%yes & yes & { }1 & { }1 & { }1
%\lastline
%\end{tabular}}
%\end{table}

\begin{table}
\caption{Front Pocket Gyroscope: Distribution of Predicted Activity Label During Test Segment on Varying Phone Orientation. \textit{These results are for the segment of the test data collection in which the phone's orientation was varied. In each subtable, the column headings show the true activity label and the row headings show the predicted activity label. There are two subcolumns under each column heading, the first for classifications based on tri-axial data and the second for classifications based on magnitude data. Each subcolumn gives the distribution of predicted activity labels, when the true activity label is that given by the column heading. Each subcolumn sums to 1. The shaded diagonal highlights the proportion of predicted activity labels that match the true activity label. The average of the diagonal elements for each data type (tri-axial vs. magnitude) is 0.51 vs. 0.80 for Participant 1, 0.51 vs. 0.45 for Participant 2, 0.55 vs. 0.62 for Participant 3, and 0.54 vs. 0.58 for Participant 4.} \label{table:vary_orientation_frontgyro}}
\centering
\scalebox{0.9}{
\small\addtolength{\tabcolsep}{-2pt}
\begin{tabular}{rllllllll}
  \hline
  & \multicolumn{8}{c}{\bf{(a) Participant 1 }} \\%(0.51, 0.80)
    \hline
  \tabularnewline
  & \multicolumn{2}{c}{stand} & \multicolumn{2}{c}{walk} & \multicolumn{2}{c}{stairUp} & \multicolumn{2}{c}{stairDown} \\
  \hline
stand & \cellcolor[gray]{0.9}{0.27} & \cellcolor[gray]{0.9}{0.49} & 0.00 & 0.00 & 0.00 & 0.00 & 0.00 & 0.00 \\ 
  walk & 0.00 & 0.00 & \cellcolor[gray]{0.9}{0.38} & \cellcolor[gray]{0.9}{0.90} & 0.00 & 0.00 & 0.25 & 0.07 \\ 
  stairUp & 0.09 & 0.00 & 0.34 & 0.09 & \cellcolor[gray]{0.9}{0.94} & \cellcolor[gray]{0.9}{1.00} & 0.32 & 0.14 \\ 
  stairDown & 0.00 & 0.10 & 0.27 & 0.01 & 0.00 & 0.00 & \cellcolor[gray]{0.9}{0.43} & \cellcolor[gray]{0.9}{0.80} \\ 
  standToSit & 0.00 & 0.07 & 0.00 & 0.00 & 0.05 & 0.00 & 0.00 & 0.00 \\ 
  sit & 0.64 & 0.27 & 0.00 & 0.00 & 0.01 & 0.00 & 0.00 & 0.00 \\ 
  sitToStand & 0.00 & 0.06 & 0.00 & 0.00 & 0.00 & 0.00 & 0.00 & 0.00 \\
   \tabularnewline
   \hline
  & \multicolumn{8}{c}{\bf{(b) Participant 2 }} \\%(0.51, 0.45)
    \hline
  \tabularnewline
  & \multicolumn{2}{c}{stand} & \multicolumn{2}{c}{walk} & \multicolumn{2}{c}{stairUp} & \multicolumn{2}{c}{stairDown} \\
  \hline
  stand & \cellcolor[gray]{0.9}{0.04} & \cellcolor[gray]{0.9}{0.03} & 0.00 & 0.00 & 0.00 & 0.00 & 0.00 & 0.00 \\ 
  walk & 0.00 & 0.00 & \cellcolor[gray]{0.9}{0.87} & \cellcolor[gray]{0.9}{0.74} & 0.14 & 0.16 & 0.57 & 0.25 \\ 
  stairUp & 0.00 & 0.39 & 0.02 & 0.21 & \cellcolor[gray]{0.9}{0.71} & \cellcolor[gray]{0.9}{0.61} & 0.00 & 0.35 \\ 
  stairDown & 0.27 & 0.01 & 0.11 & 0.05 & 0.15 & 0.23 & \cellcolor[gray]{0.9}{0.43} & \cellcolor[gray]{0.9}{0.40} \\ 
  standToSit & 0.00 & 0.15 & 0.00 & 0.00 & 0.00 & 0.00 & 0.00 & 0.00 \\ 
  sit & 0.49 & 0.38 & 0.00 & 0.00 & 0.00 & 0.00 & 0.00 & 0.00 \\ 
  sitToStand & 0.20 & 0.03 & 0.00 & 0.00 & 0.00 & 0.00 & 0.01 & 0.00 \\ 
   \tabularnewline
   \hline
  & \multicolumn{8}{c}{\bf{(c) Participant 3 }} \\%(0.55, 0.62)
    \hline
  \tabularnewline
  & \multicolumn{2}{c}{stand} & \multicolumn{2}{c}{walk} & \multicolumn{2}{c}{stairUp} & \multicolumn{2}{c}{stairDown} \\
  \hline
  stand & \cellcolor[gray]{0.9}{0.20} & \cellcolor[gray]{0.9}{0.51} & 0.00 & 0.00 & 0.00 & 0.00 & 0.00 & 0.00 \\ 
  walk & 0.00 & 0.00 & \cellcolor[gray]{0.9}{0.38} & \cellcolor[gray]{0.9}{0.80} & 0.15 & 0.28 & 0.05 & 0.27 \\ 
  stairUp & 0.01 & 0.04 & 0.24 & 0.06 & \cellcolor[gray]{0.9}{0.71} & \cellcolor[gray]{0.9}{0.67} & 0.03 & 0.22 \\ 
  stairDown & 0.04 & 0.00 & 0.38 & 0.15 & 0.14 & 0.05 & \cellcolor[gray]{0.9}{0.91} & \cellcolor[gray]{0.9}{0.51} \\ 
  standToSit & 0.00 & 0.08 & 0.00 & 0.00 & 0.00 & 0.00 & 0.00 & 0.00 \\ 
  sit & 0.76 & 0.38 & 0.00 & 0.00 & 0.00 & 0.00 & 0.01 & 0.00 \\ 
  sitToStand & 0.00 & 0.00 & 0.00 & 0.00 & 0.00 & 0.00 & 0.00 & 0.00 \\ 
   \tabularnewline
   \hline
  & \multicolumn{8}{c}{\bf{(d) Participant 4 }} \\%(0.54, 0.58)
    \hline
  \tabularnewline
  & \multicolumn{2}{c}{stand} & \multicolumn{2}{c}{walk} & \multicolumn{2}{c}{stairUp} & \multicolumn{2}{c}{stairDown} \\
  \hline
  stand & \cellcolor[gray]{0.9}{0.29} & \cellcolor[gray]{0.9}{0.38} & 0.00 & 0.00 & 0.00 & 0.00 & 0.00 & 0.00 \\ 
  walk & 0.00 & 0.02 & \cellcolor[gray]{0.9}{0.48} & \cellcolor[gray]{0.9}{0.88} & 0.30 & 0.11 & 0.17 & 0.41 \\ 
  stairUp & 0.00 & 0.00 & 0.49 & 0.12 & \cellcolor[gray]{0.9}{0.70} & \cellcolor[gray]{0.9}{0.89} & 0.12 & 0.35 \\ 
  stairDown & 0.05 & 0.02 & 0.03 & 0.01 & 0.00 & 0.00 & \cellcolor[gray]{0.9}{0.71} & \cellcolor[gray]{0.9}{0.17} \\ 
  standToSit & 0.00 & 0.00 & 0.00 & 0.00 & 0.00 & 0.00 & 0.00 & 0.04 \\ 
  sit & 0.65 & 0.56 & 0.00 & 0.00 & 0.00 & 0.00 & 0.00 & 0.00 \\ 
  sitToStand & 0.00 & 0.01 & 0.00 & 0.00 & 0.00 & 0.00 & 0.00 & 0.03 \\
  \hline
\end{tabular}}
\end{table}

\begin{table}
\caption{Front Pocket Accelerometer: Distribution of Predicted Activity Label During Test Segment on Varying Phone Orientation. \textit{See Table \ref{table:vary_orientation_frontgyro} for details. The average of the diagonal elements for each data type (tri-axial vs. magnitude) is 0.58 vs. 0.73 for Participant 1, 0.30 vs. 0.55 for Participant 2, 0.59 vs. 0.81 for Participant 3, and 0.31 vs. 0.74 for Participant 4.} \label{table:vary_orientation_frontacc}}
\centering
\scalebox{0.9}{
\small\addtolength{\tabcolsep}{-2pt}
\begin{tabular}{rllllllll}
  \hline
  & \multicolumn{8}{c}{\bf{(a) Participant 1 }} \\%(0.58, 0.73)
    \hline
  \tabularnewline
  & \multicolumn{2}{c}{stand} & \multicolumn{2}{c}{walk} & \multicolumn{2}{c}{stairUp} & \multicolumn{2}{c}{stairDown} \\
  \hline
  stand & \cellcolor[gray]{0.9}{0.94} & \cellcolor[gray]{0.9}{0.55} & 0.12 & 0.00 & 0.35 & 0.00 & 0.44 & 0.00 \\ 
  walk & 0.00 & 0.00 & \cellcolor[gray]{0.9}{0.65} & \cellcolor[gray]{0.9}{0.52} & 0.13 & 0.02 & 0.20 & 0.08 \\ 
  stairUp & 0.06 & 0.00 & 0.10 & 0.33 & \cellcolor[gray]{0.9}{0.49} & \cellcolor[gray]{0.9}{0.98} & 0.14 & 0.06 \\ 
  stairDown & 0.00 & 0.00 & 0.14 & 0.12 & 0.02 & 0.00 & \cellcolor[gray]{0.9}{0.22} & \cellcolor[gray]{0.9}{0.86} \\ 
  standToSit & 0.00 & 0.11 & 0.00 & 0.02 & 0.01 & 0.00 & 0.00 & 0.00 \\ 
  sit & 0.00 & 0.34 & 0.00 & 0.00 & 0.00 & 0.00 & 0.00 & 0.00 \\ 
  sitToStand & 0.00 & 0.00 & 0.00 & 0.00 & 0.00 & 0.00 & 0.00 & 0.00 \\ 
  \tabularnewline
    \hline
  & \multicolumn{8}{c}{\bf{(b) Participant 2 }} \\%(0.30, 0.55)
    \hline
  \tabularnewline
  & \multicolumn{2}{c}{stand} & \multicolumn{2}{c}{walk} & \multicolumn{2}{c}{stairUp} & \multicolumn{2}{c}{stairDown} \\
  \hline
  stand & \cellcolor[gray]{0.9}{0.14} & \cellcolor[gray]{0.9}{0.17} & 0.00 & 0.00 & 0.00 & 0.00 & 0.00 & 0.00 \\ 
  walk & 0.00 & 0.30 & \cellcolor[gray]{0.9}{0.24} & \cellcolor[gray]{0.9}{0.95} & 0.04 & 0.40 & 0.11 & 0.26 \\ 
  stairUp & 0.57 & 0.03 & 0.50 & 0.02 & \cellcolor[gray]{0.9}{0.54} & \cellcolor[gray]{0.9}{0.34} & 0.49 & 0.01 \\ 
  stairDown & 0.21 & 0.03 & 0.03 & 0.01 & 0.28 & 0.21 & \cellcolor[gray]{0.9}{0.28} & \cellcolor[gray]{0.9}{0.72} \\ 
  standToSit & 0.00 & 0.11 & 0.01 & 0.01 & 0.05 & 0.02 & 0.02 & 0.00 \\ 
  sit & 0.07 & 0.36 & 0.21 & 0.00 & 0.09 & 0.00 & 0.10 & 0.00 \\ 
  sitToStand & 0.00 & 0.00 & 0.00 & 0.00 & 0.00 & 0.02 & 0.00 & 0.00 \\ 
  \tabularnewline
      \hline
  & \multicolumn{8}{c}{\bf{(c) Participant 3}} \\
    \hline
  \tabularnewline
  & \multicolumn{2}{c}{stand} & \multicolumn{2}{c}{walk} & \multicolumn{2}{c}{stairUp} & \multicolumn{2}{c}{stairDown} \\
  \hline
  stand & \cellcolor[gray]{0.9}{0.96} & \cellcolor[gray]{0.9}{0.93} & 0.03 & 0.00 & 0.20 & 0.00 & 0.02 & 0.00 \\ 
  walk & 0.01 & 0.01 & \cellcolor[gray]{0.9}{0.68} & \cellcolor[gray]{0.9}{0.64} & 0.14 & 0.03 & 0.64 & 0.19 \\ 
  stairUp & 0.01 & 0.03 & 0.21 & 0.18 & \cellcolor[gray]{0.9}{0.57} & \cellcolor[gray]{0.9}{0.96} & 0.17 & 0.09 \\ 
  stairDown & 0.00 & 0.00 & 0.07 & 0.18 & 0.00 & 0.01 & \cellcolor[gray]{0.9}{0.16} & \cellcolor[gray]{0.9}{0.72} \\ 
  standToSit & 0.02 & 0.02 & 0.01 & 0.00 & 0.09 & 0.00 & 0.01 & 0.00 \\ 
  sit & 0.00 & 0.02 & 0.00 & 0.00 & 0.00 & 0.00 & 0.00 & 0.00 \\ 
  sitToStand & 0.00 & 0.00 & 0.00 & 0.00 & 0.00 & 0.00 & 0.00 & 0.00 \\
  \tabularnewline
        \hline
  & \multicolumn{8}{c}{\bf{(d) Participant 4}} \\
    \hline
  \tabularnewline
  & \multicolumn{2}{c}{stand} & \multicolumn{2}{c}{walk} & \multicolumn{2}{c}{stairUp} & \multicolumn{2}{c}{stairDown} \\
  \hline
  stand & \cellcolor[gray]{0.9}{0.00} & \cellcolor[gray]{0.9}{0.92} & 0.00 & 0.00 & 0.00 & 0.00 & 0.00 & 0.00 \\ 
  walk & 0.03 & 0.01 & \cellcolor[gray]{0.9}{0.67} & \cellcolor[gray]{0.9}{0.76} & 0.18 & 0.14 & 0.42 & 0.26 \\ 
  stairUp & 0.00 & 0.05 & 0.03 & 0.22 & \cellcolor[gray]{0.9}{0.31} & \cellcolor[gray]{0.9}{0.76} & 0.06 & 0.21 \\ 
  stairDown & 0.00 & 0.02 & 0.04 & 0.02 & 0.04 & 0.07 & \cellcolor[gray]{0.9}{0.25} & \cellcolor[gray]{0.9}{0.53} \\ 
  standToSit & 0.00 & 0.00 & 0.00 & 0.00 & 0.00 & 0.02 & 0.00 & 0.00 \\ 
  sit & 0.97 & 0.00 & 0.26 & 0.00 & 0.47 & 0.00 & 0.27 & 0.00 \\ 
  sitToStand & 0.00 & 0.01 & 0.00 & 0.00 & 0.00 & 0.01 & 0.00 & 0.00 \\
  \hline
\end{tabular}}
\end{table}

\end{document}